\documentclass[]{bytedance_seed}

\usepackage[toc,page,header]{appendix}
\usepackage{siunitx} 
\DeclareSIUnit{\kcal}{kcal}
\DeclareSIUnit{\GPUhour}{GPU\mbox{-}hour} 
\DeclareSIUnit{\e}{e}
\DeclareSIUnit{\bar}{bar}
\DeclareSIUnit{\angstrom}{\text{\AA}}
\DeclareSIUnit\Molar{\text{M}}

\usepackage{minitoc}
\usepackage{textgreek}
\usepackage{makecell}
\usepackage{float}

\title{Development and large-scale benchmarks of a protein\textendash ligand absolute binding free energy toolkit}

\author[1,*]{Ye Ding}
\author[1,*]{Yu Liu}
\author[1,*]{Ailun Wang}
\author[1]{Yu Xia}
\author[1,\dagger]{Zhi Wang}
\author[1,\dagger]{Wen Yan}

\affiliation[1]{ByteDance Seed}

\contribution[*]{Equal contribution}
\contribution[\dagger]{Corresponding authors}

\abstract{
    Absolute binding free energy (ABFE) calculations offer a theoretically rigorous approach for predicting protein--ligand binding affinities without the scaffold constraints of relative binding free energy (RBFE) perturbations.
    However, broad adoption of ABFE in high-throughput hit discovery campaigns has been hindered by high computational costs and a lack of large-scale validation.
    Here, we present Felis, a source-available, automated, and scalable ABFE toolkit that integrates RelocationMC to enhance binding-site water sampling.
    Using the ByteFF25-AM1-BCC force field, ABFE calculations with Felis yielded similar average rank correlations to FEP+ RBFE across 43 benchmark systems comprising 857 ligand entries.
    The offset-adjusted errors were larger than those of FEP+ RBFE.
    In scytalone dehydratase, RelocationMC rapidly hydrated an initially dry cavity and reduced the dependence of binding free energy offsets on the initial water configuration.
    Across the benchmark, it reduced absolute offsets in 27 of 43 systems without a loss of aggregate predictive performance.
    We also evaluated sampling-length dependence and ligand ranking for a challenging KRAS(G12D) dataset containing highly charged ligands and a charged cofactor.
    Neither force-field parameters nor alchemical schedules were adjusted using experimental affinities from the benchmarks.
    These results support the use of Felis for ligand prioritization and highlight remaining sampling limitations.
}

\date{\today}
\correspondence{Zhi Wang at \email{zhi.wang1@bytedance.com}, Wen Yan at \email{wen.yan@bytedance.com}}

\begin{document}
\maketitle

\section{Introduction}

Accurate and reliable prediction of protein--ligand binding free energies is central to computer-aided drug discovery, where robust affinity ranking guides both hit discovery and lead optimization.
Data-driven approaches such as Boltz~\cite{passaroBoltz2AccurateEfficient2025,thalerBoltzABFEFreeEnergy2025,furuiBoltzinaEfficientAccurate2025} and alternative scoring methods~\cite{pecinaSQM220SemiempiricalQuantummechanical2024,gilsonRapidAccurateRanking2024} are attracting increasing interest due to their speed and scalability.
Yet physics-based free energy sampling methods such as free energy perturbation (FEP) provide a rigorous route to binding thermodynamics grounded in an explicit potential energy surface.

The ecosystem around FEP, including relative binding free energy (RBFE) and absolute binding free energy (ABFE) methods, has expanded rapidly alongside emerging large-scale benchmark datasets~\cite{rossMaximalCurrentFEPBenchmark2023,hahnCurrentStateOpen2024,baumannLargescaleCollaborativeAssessment2025,zouBreakingBarriersFEP2025}.
Prominent frameworks include FEP+~\cite{wangAchievingHighAccuracy2012,wangAccurateReliablePrediction2015,wangAccurateModelingScaffold2017,harderOPLS3ForceField2016,roosOPLS3eExtendingForce2019,luOPLS4ImprovingForce2021,dammOPLS5AdditionPolarizability2024}, pmx~\cite{gapsysPmxAutomatedProtein2015,gapsysLargeScaleRelative2020,gapsysAccurateAbsoluteFree2021,gapsysPreExascaleComputingProtein2022}, OpenFE~\cite{baumannBroadeningScopeBinding2023,baumannLargescaleCollaborativeAssessment2025}, ATM~\cite{wuAlchemicalTransferApproach2021,azimiRelativeBindingFree2022,azimiBindingSelectivityAnalysis2024,chenPerformanceAnalysisAlchemical2024,gallicchioRelativeBindingFree2025,azimiPotentialDistributionTheory2025}, and Uni-FEP~\cite{zouBreakingBarriersFEP2025}.
Specialized tools such as BAT~\cite{heinzelmannAutomationAbsoluteProteinligand2021,heinzelmannBAT2OpenSourceTool2024,heinzelmannRelativeBATAutomated2025}, QligFEP~\cite{jespersQligFEPAutomatedWorkflow2019}, AQFEP~\cite{crivelli-deckerMachineLearningGuided2024}, FEP-SPell-ABFE~\cite{liFEPSpellABFE2025}, SPONGE-FEP~\cite{xiaSPONGEFEPAutomatedRelative2025}, PyAutoFEP~\cite{carvalhomartinsPyAutoFEPAutomatedFree2021}, ALCHEMD~\cite{liWeightedCCRBFE2023,yaoCAR2024,liuALCHEMDBridgingAccessibility2025,liuStateFunctionCorrection2025}, RED-E ABFE~\cite{liuABFEREDE2023,huangRestraintABFE2025,liuDivideConquerABFE2025}, and many others have also emerged.

In practice, RBFE calculations are the dominant production workflow in computational drug discovery.
RBFE estimates the binding free energy difference ($\Delta\Delta G$) between two congeneric ligands in the same binding site.
These calculations are typically implemented by alchemically transforming one ligand into another along a thermodynamic cycle through a series of nonphysical intermediate states.
In a large benchmark study, RBFE achieved an accuracy of about \SI{1}{\kcal\per\mole} (the so-called ``chemical accuracy'') for pairwise free energy differences between ligands, compared with an experimental uncertainty of \SI{0.67}{\kcal\per\mole}~\cite{rossMaximalCurrentFEPBenchmark2023}.
RBFE is generally most reliable when ligands share a common scaffold, which helps maintain phase-space overlap and smooth topological changes.
When scaffold similarity is low (e.g., scaffold hopping), convergence may require more complex protocols, such as intermediate steps and soft-bond potentials.
Methods that broaden the scope have been proposed~\cite{baumannBroadeningScopeBinding2023,TsaiTandemFEP2025}, but practical complexities such as atom mapping and 3D alignment remain.

\begin{figure}[htb]
    \centering
    \includegraphics[width=0.5\linewidth]{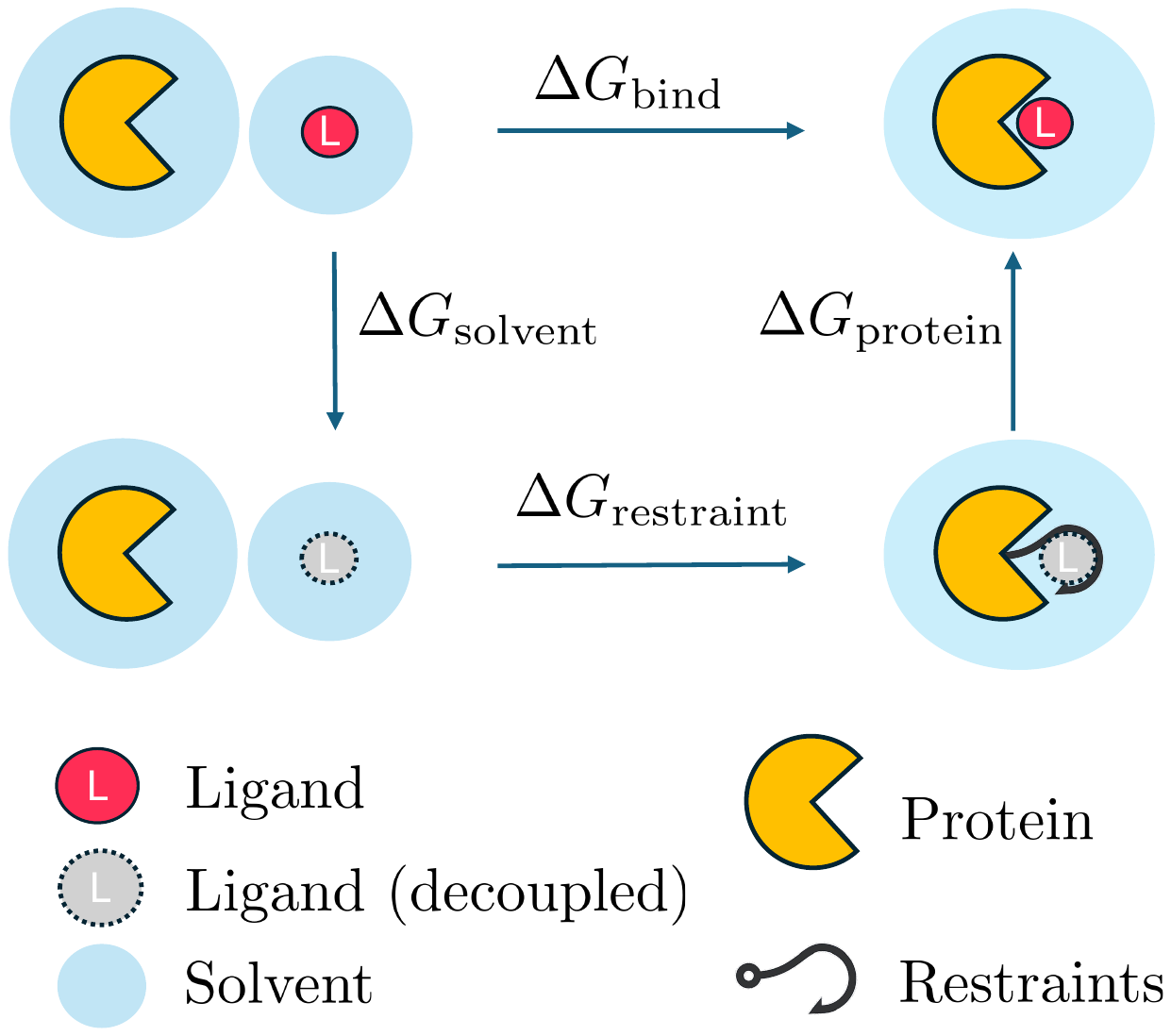}
    \caption{
    Overview of the ABFE thermodynamic cycle.
    The absolute binding free energy is given by $\Delta G_\mathrm{bind} = \Delta G_\mathrm{solvent} + \Delta G_\mathrm{restraint} + \Delta G_\mathrm{protein}$, where $\Delta G_\mathrm{solvent}$ is the free energy of decoupling the ligand from solvent, $\Delta G_\mathrm{restraint}$ is the analytical correction for the Boresch-style restraints, and $\Delta G_\mathrm{protein}$ is the free energy of coupling the ligand to the protein environment while releasing the Boresch-style restraints.
    Felis implements this cycle using a double-annihilation path that also scales intramolecular ligand nonbonded interactions.
    }
    \label{fig:abfe_workflow}
\end{figure}

ABFE avoids the scaffold-similarity assumption by scoring each ligand independently.
Like RBFE, ABFE uses a thermodynamic cycle in which a ligand is alchemically decoupled in solution and coupled in the protein binding site.
The binding free energy is obtained by summing free energy differences across alchemical intermediate states (Fig.~\ref{fig:abfe_workflow}).
This makes ABFE attractive, in principle, for early-stage hit discovery.
However, routine high-throughput ABFE remains challenging because it requires substantial sampling, particularly in the protein--ligand leg.
Water exchange is a particularly important source of sampling error because buried hydration sites can equilibrate more slowly than routine simulation timescales.
Ligand modifications can alter binding-site hydration, making free energy estimates sensitive to incomplete water exchange and the initial placement of crystallographic waters~\cite{rossEnhancingWaterSampling2020,wahlAssessingPredictivePower2019,wagleLeveragingSeparationStates2024}.
Enhanced sampling methods can mitigate this problem by accelerating water exchange between buried sites and bulk solvent~\cite{dingImprovingBindingFree2026}.

Routine ABFE also demands robust protocol choices, including restraint definitions and practical alchemical transformation schedules ($\lambda$ schedules).
Despite the growing ecosystem, gaps remain in deploying ABFE as a routine production workflow.
Real-world applications require fully automated pipelines, efficient hardware utilization, and robust diagnostics with minimal manual intervention.
Crucially, most published ABFE validation studies are much smaller than their RBFE counterparts.
For example, published benchmark results for Schr\"odinger's FEP+ ABFE~\cite{chenEnhancingHitDiscovery2023} cover only eight protein targets, far fewer than the comprehensive FEP+ RBFE validation~\cite{rossMaximalCurrentFEPBenchmark2023}.
Available benchmarks for other ABFE frameworks cover even fewer protein targets~\cite{liFEPSpellABFE2025,gapsysAccurateAbsoluteFree2021}.
This lack of extensive benchmarking hinders adoption of ABFE in production pipelines.

To address these challenges, we developed Felis (Free Energy of Ligand-protein InteractionS), an automated toolkit for ABFE calculations.
Felis combines automated system preparation and restraint selection with GPU-accelerated Hamiltonian replica-exchange molecular dynamics (HREMD) in \texttt{OpenMM}.
The CUDA-enabled \texttt{PerturbNonbondedPlugin} package reduces the overhead of evaluating neighboring alchemical states and applying replica exchanges without reinitializing simulation contexts.
Hydrogen mass repartitioning (HMR) supports a \SI{4}{\femto\second} production time step.
Felis also integrates RelocationMC, a fixed-particle Monte Carlo protocol that accelerates water exchange between buried binding sites and bulk solvent.

We benchmarked ABFE calculations with Felis on 43 benchmark systems comprising 857 ligand entries, selected from a comprehensive RBFE dataset~\cite{rossMaximalCurrentFEPBenchmark2023}.
The calculations used AMBER ff14SB for proteins and ByteFF25 for ligands and cofactors.
ByteFF25 extends the bonded-parameter training of ByteFF24~\cite{zhengByteFF24} using a larger quantum-chemistry dataset, while its AM1-BCC variant uses GAFF2-derived nonbonded parameters~\cite{wangDevelopmentTestingGeneral2004}.
We compared ByteFF25-AM1-BCC with FEP+ RBFE and evaluated the alternative ByteFF25-ABCG2 charge-model variant on the same benchmark systems.
The Felis ABFE predictions showed similar average rank correlations to FEP+ RBFE, although their offset-adjusted errors were larger.
Representative simulation timings characterize the computational cost of the ABFE workflow.
We also examined water sampling in scytalone dehydratase and assessed sampling-length dependence and ligand ranking in a challenging KRAS(G12D) series containing large, highly charged ligands.
Neither force-field parameters nor alchemical schedules were adjusted using experimental affinities from the benchmarks.

\FloatBarrier
\section{Method}

Felis uses a double-annihilation protocol for ABFE calculations.
The protocol scales ligand--environment and standard intramolecular ligand nonbonded interactions in both alchemical legs, while retaining nonzero 1--4 terms.
Initial solute structures are placed in an orthorhombic periodic box and solvated without changing their internal coordinates.
In this study, both systems were minimized using OpenMM L-BFGS, with gradual heating applied to the ligand--solvent system.
The preparation method can be configured separately for each system.
For protein--ligand systems, we include a set of Boresch-style restraints~\cite{boreschAbsoluteBindingFreeEnergies2003,wangAbsoluteBindingFreeEnergyCalculations2006,hugginsComparingThePerformance2022,chenEnhancingHitDiscovery2023,boreschAnalyticalCorrectionsRestraints2024,liFEPSpellABFE2025,huangRestraintABFE2025} between the ligand and the protein.
We run an unrestrained molecular dynamics (MD) simulation in the isothermal--isobaric (NPT) ensemble and use the resulting trajectory to determine the anchor atoms and their reference geometries.
The selected anchors define the restraints used during ABFE sampling.
After the ABFE simulations for the ligand--solvent and protein--ligand systems are complete, we post-process the results using a free-energy estimator and report the absolute binding free energy.

The restraints, alchemical transformation, water sampling strategy, and simulation protocol are described in the following subsections.
Additional methodological details are provided in the appendix.

\subsection{Restraints}

For protein--ligand alchemical simulations, we apply geometric restraints to the protein--ligand complex in the decoupled state and release them in the physical bound state.
We define these restraints using three protein anchor atoms (P1, P2, P3) and three ligand anchor atoms (L1, L2, L3).
The restraint set comprises one distance term (P1-L1), two angle terms (P1-L1-L2 and L1-P1-P2), and three dihedral terms (P2-P1-L1-L2, P1-L1-L2-L3, and L1-P1-P2-P3).
The distance and angle restraints employ a harmonic potential of the form $U(g) = (k/2)(g-g_0)^2$, where $g$ and $g_0$ represent the instantaneous and equilibrium geometries, respectively, and $k$ is the force constant.
For the dihedral restraints, we use an alternative potential, $U(g) = k[1-\cos(g-g_0)]$.
This functional form approximates the harmonic potential to second order and remains periodic across angular boundaries.
The free energy contribution of these restraints in the decoupled state is computed analytically using the closed-form expression derived by~\citet{chenEnhancingHitDiscovery2023}.
The free energy cost associated with releasing the restraints in the physical bound state is determined via free energy simulations.

Anchor atoms are selected based on an initial unrestrained MD simulation in the NPT ensemble.
The resulting trajectory is processed using the DSSP~\cite{DSSP4ProSci2025} and ProLIF~\cite{prolif2021} packages.
DSSP is employed to identify and categorize secondary structure elements for each residue.
We prioritize protein residues that maintain stable secondary structures, specifically \textalpha-helices, 3\textsubscript{10}-helices, \textpi-helices, extended strands, or \textbeta-bridges, in more than 50\% of trajectory frames.
ProLIF is used to characterize specific protein--ligand interactions.
Candidate interactions are grouped by ligand anchor atom, protein residue, and interaction type, and retained if they occur in at least 40\% of the analyzed frames.
For each pair of a ligand anchor atom and a protein residue, retained interaction types are merged by taking the union of their frames and assigning the highest interaction priority.

Protein anchors are drawn from the N, C\textalpha, and C backbone atoms of a single candidate residue, with all six orderings evaluated.
The first ligand anchor, L1, is selected from the interacting atom or atoms within three bonds of it.
Eligible L1 atoms are heavy atoms with at least two heavy-atom neighbors, excluding sulfonyl sulfur and oxygen atoms and carboxyl oxygen atoms.
L2 is a nonterminal heavy atom bonded to L1, and L3 is a distinct nonterminal heavy atom bonded to L2.
All eligible L1--L2--L3 triples are evaluated.
Candidates are retained only if the L1-L2-L3 and P1-L1-L2 angles each lie within \SIrange{45}{135}{\degree} in at least 40\% of all analyzed frames.
These qualifying frames are counted within the merged interaction frames for the candidate.
Candidates are ranked by the frequency with which the P1-L1-L2 and L1-P1-P2 angles lie in this range, with penalties for larger mean deviations from \SI{90}{\degree}.
The score also favors protein residues with stable secondary structure.
A further bonus is applied to residues with at least two retained ionic, cation--\textpi, hydrogen-bond, or halogen-bond interactions.
This bonus requires the second-most-frequent interaction to occur more than 60\% as often as the most frequent interaction.
Interaction priorities decrease from ionic interactions to cation--\textpi\ interactions, hydrogen bonds, halogen bonds, and aromatic stacking interactions.
Ties are resolved by sorting based on atom indices.

If no interaction candidates survive filtering, or if they yield no valid anchor set, the algorithm falls back to a distance-based method.
This method first selects eligible ligand heavy atoms, preferring nitrogen and oxygen when available.
For each frame, the closest pair of a selected ligand atom and a protein backbone atom is identified using minimum-image distances, with a \SI{1.0}{\nano\meter} cutoff.
Contacts are grouped by protein residue, and the most frequently associated ligand atom defines L1 for each residue.
Candidate residues must account for at least 15\% of the analyzed frames.
Anchor triples undergo the same geometric filtering, using a 15\% frame threshold.
They are ranked by the combined counts of frames satisfying the P1-L1-L2 and L1-P1-P2 angle ranges, followed by the mean angular deviations from \SI{90}{\degree}.
Frame-count thresholds are rounded down to integers in both selection methods.
If the distance-based method also yields no valid anchor set, the calculation stops with an error.

\subsection{Double-annihilation protocol}

We use this double-annihilation protocol for all absolute binding free energy calculations.
For each atom $i$, the charge at a given alchemical state is
\begin{equation}
q_i = \lambda_{\mathrm{q},A}q_{i,A} + \lambda_{\mathrm{q},B}q_{i,B}.
\end{equation}
For ligand atoms, $q_{i,A}$ is the physical partial charge and $q_{i,B}=0$.
The ABFE schedules use $\lambda_{\mathrm{q},B}=1-\lambda_{\mathrm{q},A}$, which linearly removes the ligand charges.

For van der Waals (vdW) interactions, the standard 12-6 Lennard-Jones potential is modified using the Beutler~\cite{beutlerSoftcoreLJ1994} softcore form:
\begin{equation}
U_{ij}^{\mathrm{vdW}} =
\sum_{X\in\{A,B\}}
4\lambda_{\mathrm{vdW},X}\epsilon_{ij,X}
\left[
\left(\frac{\sigma_{ij,X}^{6}}{r_{\mathrm{eff},X}^{6}}\right)^2
-\frac{\sigma_{ij,X}^{6}}{r_{\mathrm{eff},X}^{6}}
\right],
\end{equation}
where
\begin{equation}
r_{\mathrm{eff},X}^{6}
= r_{ij}^{6}+\alpha\left(1-\lambda_{\mathrm{vdW},X}\right)^2\sigma_{ij,X}^{6}.
\end{equation}
The pair parameters follow the Lorentz--Berthelot mixing rules, $\sigma_{ij,X}=(\sigma_{i,X}+\sigma_{j,X})/2$ and $\epsilon_{ij,X}=\sqrt{\epsilon_{i,X}\epsilon_{j,X}}$.
For ligand atoms, the physical A-state parameters are retained, whereas $\epsilon_{i,B}=0$ and $\sigma_{i,B}=\sigma_{i,A}$.
The schedules use $\lambda_{\mathrm{vdW},B}=1-\lambda_{\mathrm{vdW},A}$, and the softcore parameter is $\alpha=1$.
The softcore form is applied to every non-exception vdW pair that contains at least one ligand atom, including intramolecular ligand pairs.
Here, non-exception pairs are evaluated using the standard mixing rules rather than explicitly assigned pair parameters or exclusions.
Nonzero 1--4 Coulomb and vdW interactions are evaluated separately using their original parameters and are not alchemically scaled.

Periodic torsional terms associated with rotatable ligand bonds are scaled by a separate torsional alchemical parameter.
Rotatable bonds are identified using the Torsion Fingerprints algorithm as implemented in \texttt{RDKit}~\cite{RDKitOSS}.
Other bonded terms remain fully coupled.

Felis evaluates the alchemical nonbonded interactions with the \texttt{PerturbNonbondedForce} class provided by the CUDA-enabled \texttt{PerturbNonbondedPlugin} package.
The plugin was developed by adapting the \texttt{OpenMM NonbondedForce} implementation.
The plugin modifies the CUDA kernels and parameter-passing mechanism to store the complete alchemical schedule and select the active state through a global \texttt{replica\_id} context parameter.
Changing this parameter updates the charge and vdW parameters at the next force evaluation without reinitializing the \texttt{OpenMM} context.
This design reduces the overhead of evaluating neighboring Hamiltonians and applying accepted HREMD exchanges.
This makes frequent exchange attempts practical, while mixing efficiency still depends on the overlap between neighboring states~\cite{sindhikaraExchangeOftenProperly2010}.

Electrostatic neutrality is maintained in all simulation systems.
In the physical state, the net charges are neutralized by the addition of sodium or chloride counterions.
For charged ligands, selected solvent molecules are designated as ``alchemical waters''~\cite{chenAlchemicalWater2018}.
Their A-state charges are the standard water-model charges, whereas their B-state charges carry the net charge removed from the ligand.
Positive compensating charge is assigned to the oxygen atom, whereas negative compensating charge is divided equally between the two hydrogen atoms.
The number of alchemical waters is chosen so that the charge change assigned to each water lies within \SIrange{-0.4}{0.6}{\e}.
Positive compensation increases the oxygen charge by at most \SI{0.6}{\e} per water, whereas negative compensation reduces each hydrogen charge by at most \SI{0.2}{\e}.
With the TIP3P charges, these limits keep the oxygen charge negative and the hydrogen charges positive in both end states.
Their charges follow the same complementary A-to-B interpolation as the ligand, which preserves the total system charge in every thermodynamic state.

\subsection{RelocationMC water-sampling protocol}

Water exchange between buried binding sites and bulk solvent can be slower than the timescales sampled in routine binding free energy calculations.
Several methods have been developed to address this limitation.
Grand canonical Monte Carlo (GCMC) inserts and deletes water molecules at a specified chemical potential~\cite{rossEnhancingWaterSampling2020}, whereas nonequilibrium candidate Monte Carlo can improve the acceptance of these grand canonical moves~\cite{mellingEnhancedGrandCanonical2023}.
Fixed-particle alternatives combine molecular dynamics with translational Monte Carlo moves that exchange water molecules between buried sites and bulk solvent~\cite{benShalomSimulatingWater2019,benShalomFastEquilibration2021}.
WaterMap and grid inhomogeneous solvation theory instead characterize the spatial thermodynamics of hydration sites from explicit-solvent trajectories~\cite{wangLigandBindingWetDry2011,nguyenGridInhomogeneousSolvation2012}.
Specialized alchemical treatments have also been developed for buried or slowly exchanging waters in binding free energy calculations~\cite{geAbsoluteBindingFree2022,liuDivideConquerABFE2025}.

RelocationMC is implemented in the CUDA-enabled OpenMM package \texttt{RelocationMCPlugin} and refines the fixed-particle Swap Monte Carlo approach~\cite{dingImprovingBindingFree2026}.
It relocates complete water molecules between a ligand-centered inner sphere and bulk solvent while retaining a constant number of particles in the NPT ensemble.
This avoids particle insertion or deletion and calibration of a water chemical potential.

RelocationMC uses a paired-pool multiple-try Metropolis (MTM) proposal~\cite{liuMultipleTryMethod2000}.
Source selection favors waters with less favorable interactions with their environment.
Trial poses combine uniformly sampled whole-box positions with independent Haar-distributed rigid-body rotations.
After region classification, a candidate in the opposite region is selected using Boltzmann weights based on its interaction with the environment after removal of the source water.
The reverse pool retains the unselected candidates and replaces the selected candidate with the original water pose.
The acceptance rule accounts for the relocation direction, source and candidate selection probabilities, and the full potential-energy change to satisfy detailed balance for the implemented finite-grid particle-mesh Ewald (PME) Hamiltonian.

The CUDA implementation evaluates candidate direct-space and reciprocal-space PME interactions in parallel, including the finite-grid correction required for acceptance.
Accepted moves preserve the water's center-of-mass velocity and rotate its internal velocities consistently with its coordinates.
The proposal construction, acceptance-ratio derivation, PME energy decomposition, velocity update, and production settings are detailed in Appendix~\ref{app:relocationmc-details}.

\subsection{Simulation protocol}

All ABFE production simulations are performed with \texttt{OpenMM} 8.5 on the CUDA platform using mixed precision.
The equations of motion are propagated with the \texttt{LangevinMiddleIntegrator}~\cite{zhangLangevinMiddle2019} at \SI{298.15}{\kelvin} and a friction coefficient of \SI{0.5}{\per\pico\second}.
Hydrogen masses are repartitioned to \SI{1.5}{\dalton}~\cite{feenstraHydrogenMass1999,hopkinsLongTimeStep2015,jungOptimizedHydrogenMass2021}.
Bonds involving hydrogen are constrained, and water molecules are kept rigid.
These settings permit a \SI{4}{\femto\second} integration time step.
Production simulations are conducted in the NPT ensemble at \SI{1.01325}{\bar}.
Monte Carlo barostat moves are attempted every 25 integration steps (\SI{0.1}{\pico\second}).
Electrostatic interactions are evaluated with PME using a \SI{1.0}{\nano\meter} real-space cutoff and an Ewald error tolerance of $5\times10^{-4}$.
The van der Waals cutoff is also \SI{1.0}{\nano\meter}.
During alchemical production, \texttt{PerturbNonbondedForce} applies a lambda-dependent analytical long-range dispersion correction.
The systems use the TIP3P water model~\cite{jorgensen1983comparison} and the AMBER ff14SB protein force field~\cite{Maierff14sb2015}.
Na$^+$ and Cl$^-$ counterions use the TIP3P-specific Joung--Cheatham parameters~\cite{joung2008determination,joung2009molecular}.
Mg$^{2+}$ uses the Li--Merz 12--6 parameters~\cite{LiMerzIonParameters2015} supplied in the AMBER TIP3P ion parameter set.

For the ligand--solvent system, default relaxation begins with up to \num{1000} iterations of OpenMM L-BFGS minimization.
It is then heated from \SI{10}{\kelvin} to \SI{300}{\kelvin} in \SI{10}{\kelvin} increments, with \num{10000} integration steps at each temperature.

For the protein--ligand system, we used OpenMM L-BFGS minimization before NPT sampling.
Boresch restraint anchors are selected from four concurrent NPT trajectories at the target temperature.
Each trajectory is \SI{3}{\nano\second} long, with frames saved every \SI{10}{\pico\second}.
The first \SI{1}{\nano\second} of each trajectory is discarded, and the remaining frames are combined for anchor selection.

The alchemical calculations use Hamiltonian replica-exchange molecular dynamics (HREMD) in the ABFE v2 workflow of Felis v0.2.0.
Here, v0.2.0 identifies the software release, whereas ABFE v2 identifies the workflow.
For neutral ligands, the default schedules contain 24 states in the ligand--solvent leg and 80 in the protein--ligand leg.
For charged ligands, they contain 32 states in the ligand--solvent leg and 96 in the protein--ligand leg.
A user-supplied schedule can replace these defaults.
Unless otherwise specified, each thermodynamic state is propagated for \SI{5}{\nano\second}.
For these benchmark analyses, all recorded production work samples within each interval are retained, with no additional equilibration discard.
Neighbor exchanges are attempted every 250 integration steps (\SI{1}{\pico\second}) using alternating even--odd pairs.
At each exchange point, every worker evaluates its configuration under its current and neighboring Hamiltonians, including the $PV$ contribution.
A coordinator then applies the Metropolis criterion.
For an accepted exchange, each worker updates its \texttt{replica\_id} context parameter, which selects the corresponding alchemical $\lambda$ values.
Coordinates remain on the same worker and are not transferred.
Trajectory frames and restart checkpoints are written every \num{125000} steps (\SI{0.5}{\nano\second}).

Felis creates one \texttt{OpenMM} context and one worker process per thermodynamic state.
Contiguous groups of workers are distributed as evenly as possible across the visible GPUs.
Multiple contexts on each GPU execute concurrently through the CUDA Multi-Process Service.
Exchange decisions are coordinated across the full thermodynamic-state ladder, including neighboring states assigned to different GPUs.
This design avoids peer-to-peer coordinate transfers while retaining global synchronization for HREMD exchanges and checkpointing.

Enhanced water sampling in the protein--ligand leg uses the CUDA-enabled OpenMM \texttt{RelocationMCPlugin} package.
RelocationMC proposes rigid-body moves of complete water molecules between a ligand-centered inner sphere and the bulk solvent.
These moves alter the water occupancy near the binding site while preserving the total number of water molecules.
The sphere is centered on the ligand heavy atoms.
Its fixed radius is one-half of the maximum minimum-image separation between ligand heavy atoms in the minimized bound pose, plus \SI{0.3}{\nano\meter}.
A relocation is attempted every \num{1000} integration steps (\SI{4}{\pico\second}).
Each attempt uses a symmetric paired-pool multiple-try Metropolis protocol with $K=16384$ candidate poses.
Each candidate combines an independently sampled anchor with a Haar-distributed rigid-body orientation.
The acceptance probability accounts for the full potential-energy change, including direct-space and reciprocal-space PME contributions and the finite-grid correction.
It also includes the source-selection and multiple-try proposal weights.
For charged ligands, the alchemical waters used for charge compensation are excluded from the set of relocatable molecules.

For each alchemical leg, reduced work values are constructed from energy evaluations at neighboring states.
The free-energy increment for each neighboring pair is estimated with the Bennett acceptance ratio (BAR)~\cite{bennettBAR1976}, using an implementation derived from PyMBAR~\cite{shirtsMBAR2008}.
These increments are summed along the alchemical path.
Their statistical uncertainties are combined in quadrature.
The adjacent-state calculations also provide overlap and block-convergence diagnostics.
These diagnostics are included in the accompanying raw data for all ten independent MRTX1133 calculations at each sampling duration.
The absolute binding free energy combines the ligand--solvent and protein--ligand leg free energies with the analytical Boresch restraint correction.

\FloatBarrier
\section{Benchmarks}

\subsection{Dataset}
To benchmark ABFE predictions from Felis across a diverse set of protein targets and ligands, we used the Schr\"odinger 2023 benchmark set compiled by~\citet{rossMaximalCurrentFEPBenchmark2023}.
Structure files, experimental affinity measurements, and FEP+ RBFE predictions were obtained from the public repository\footnote{\url{https://github.com/schrodinger/public_binding_free_energy_benchmark}}.
Because this benchmark was originally designed for RBFE, some systems contain only a small number of ligand entries.
The retained subset has the following ligand-coverage, structural, and experimental-affinity characteristics:
\begin{enumerate}
    \item \textbf{Sufficient ligand coverage} Each retained benchmark system contains at least six ligand entries.
    Variants of the same ligand (e.g., multiple protonation states or conformations) are generally combined into one ligand entry, except where the original benchmark assignment treats them as distinct entries.
    \item \textbf{Clean binding sites} Systems with cofactors or metal ions that interact directly with the ligand near the binding site were excluded.
    \item \textbf{Compatible with AMBER ff14SB} Targets containing non-canonical residues unsupported by the AMBER ff14SB force field~\cite{Maierff14sb2015} were excluded.
    \item \textbf{Experimental affinity coverage} Experimental binding free energies from the public benchmark were used as reference values.
    All retained reference values are negative, and each system contains at least six distinct experimental affinity levels.
    For this dataset description, values differing by less than \SI{1e-5}{\kcal\per\mole} are counted as a single affinity level.
    Tied experimental measurements were retained, with ties accounted for in the rank statistics.
\end{enumerate}
The retained subset comprises 43 benchmark systems and 857 ligand entries.
Table~\ref{tab:schrodinger_benchmark_sets} summarizes the counts for each curated subset.
This selected subset is comparable in size to the ``public benchmark set'' in the recent OpenFE collaborative assessment~\cite{baumannLargescaleCollaborativeAssessment2025}.

We made minimal adjustments to the target proteins to ensure workflow compatibility and simulation stability:
\begin{enumerate}
    \item Missing terminal capping atoms were added (hydrogen or acetyl at the N-terminus, oxygen or N-methylamide at the C-terminus).
    \item Crystallographic water molecules and ions were retained as provided in the original repository.
    Water molecules severely clashing with the protein or ligand were removed.
    \item Protonation states of both protein structures and ligands remained unchanged from the original structures.
\end{enumerate}

To prepare for Felis ABFE simulations, each protein structure was parameterized using the AMBER ff14SB force field.
Ligands and cofactor molecules were parameterized using the in-house ByteFF25 force field, which builds on ByteFF24~\cite{zhengByteFF24}.
Force-field development was independent of the benchmark, and neither force-field parameters nor alchemical schedules were tuned using its experimental affinities.

\begin{table}[t]
\centering
\caption{Curated subsets selected from the Schr\"odinger FEP+ public benchmark repository.
Counts refer to the retained benchmark subset described in the main text.
}
\label{tab:schrodinger_benchmark_sets}
\begin{tabular}{lrr}
\toprule
Subset & No. of systems & No. of ligand entries \\
\midrule
jacs          & 8 & 199 \\
merck         & 8 & 262 \\
gpcrs         & 4 & 100  \\
waterset      & 5 & 47  \\
misc          & 3 & 60  \\
opls\_stress   & 4 & 55  \\
fragments     & 6 & 55  \\
macrocycles   & 2 & 17  \\
janssen\_bace  & 3 & 62  \\
\bottomrule
\end{tabular}
\end{table}

\subsection{Simulation performance}
\label{sec:simulation_performance}

We assessed computational cost using a representative PTP1B protein--ligand system containing \num{42034} atoms.
The timing configuration used 96 thermodynamic states distributed across eight GPUs, with 12 replicas per GPU, hydrogen mass repartitioning, and a \SI{4}{\femto\second} time step.
Neighbor exchanges were attempted every 250 integration steps (\SI{1}{\pico\second}).
Propagating each state for \SI{5}{\nano\second} corresponds to an aggregate sampling duration of \SI{480}{\nano\second} across the 96 replicas.
We define GPU time as elapsed wall time multiplied by the number of allocated GPUs.

The measured wall time for \SI{1}{\nano\second} per state was \SI{13.2}{\minute} for production alone and approximately \SI{14}{\minute} including initialization.
The production rate corresponds to approximately \SI{109}{\nano\second\per\day} per replica.
Extrapolating the production rate to \SI{5}{\nano\second} per state gives \SI{66}{\minute} of wall time and \SI{8.8}{\GPUhour} (Table~\ref{tab:simulation_performance}).
Scaling the initialization-inclusive \SI{1}{\nano\second} timing gives a conservative estimate of \SI{70}{\minute} and \SI{9.3}{\GPUhour}.
Because initialization is a one-time cost, the latter is a rate-based estimate rather than a direct measurement of a \SI{5}{\nano\second} calculation.

\begin{table}[H]
\centering
\caption{
Representative simulation timings for the 96-state PTP1B protein--ligand leg on eight GPUs with HMR enabled and RelocationMC disabled.
Measurements cover \SI{1}{\nano\second} per state; the \SI{5}{\nano\second} values are rate-based extrapolations.
The initialization-inclusive estimate scales the startup contribution with the measured rate.
}
\label{tab:simulation_performance}
\begin{tabular}{lrrr}
\toprule
Timing basis & \makecell{Measured \SI{1}{\nano\second}\\wall time (min)} & \makecell{Estimated \SI{5}{\nano\second}\\wall time (min)} & \makecell{Estimated cost\\(GPU-hours)} \\
\midrule
Production only & 13.2 & 66 & 8.8 \\
Initialization-inclusive rate & 14 & 70 & 9.3 \\
\bottomrule
\end{tabular}
\end{table}

For the same PTP1B calculation, the ligand--solvent leg completed in less than \SI{15}{\minute} on eight GPUs, corresponding to less than \SI{2}{\GPUhour}.
System preparation and postprocessing together required less than \SI{10}{\minute} of elapsed time on the CPU.
Sampling in the protein--ligand leg therefore dominated the GPU cost of this example.

\subsection{Case study of water sampling}
Slow exchange between buried hydration sites and bulk solvent can impair convergence and make binding free energy estimates sensitive to the initial water configuration~\cite{rossEnhancingWaterSampling2020,dingImprovingBindingFree2026}.
Felis addresses this problem with RelocationMC, which moves water molecules between a ligand-centered region and bulk solvent while preserving the total number of particles.
We evaluated this protocol using scytalone dehydratase, a water-sensitive system with a deeply buried hydration site~\cite{woodSystematicErrorsFree1991,wagleLeveragingSeparationStates2024,wahlAssessingPredictivePower2019}.

\begin{figure}[t]
\centering
\includegraphics[width=\linewidth]{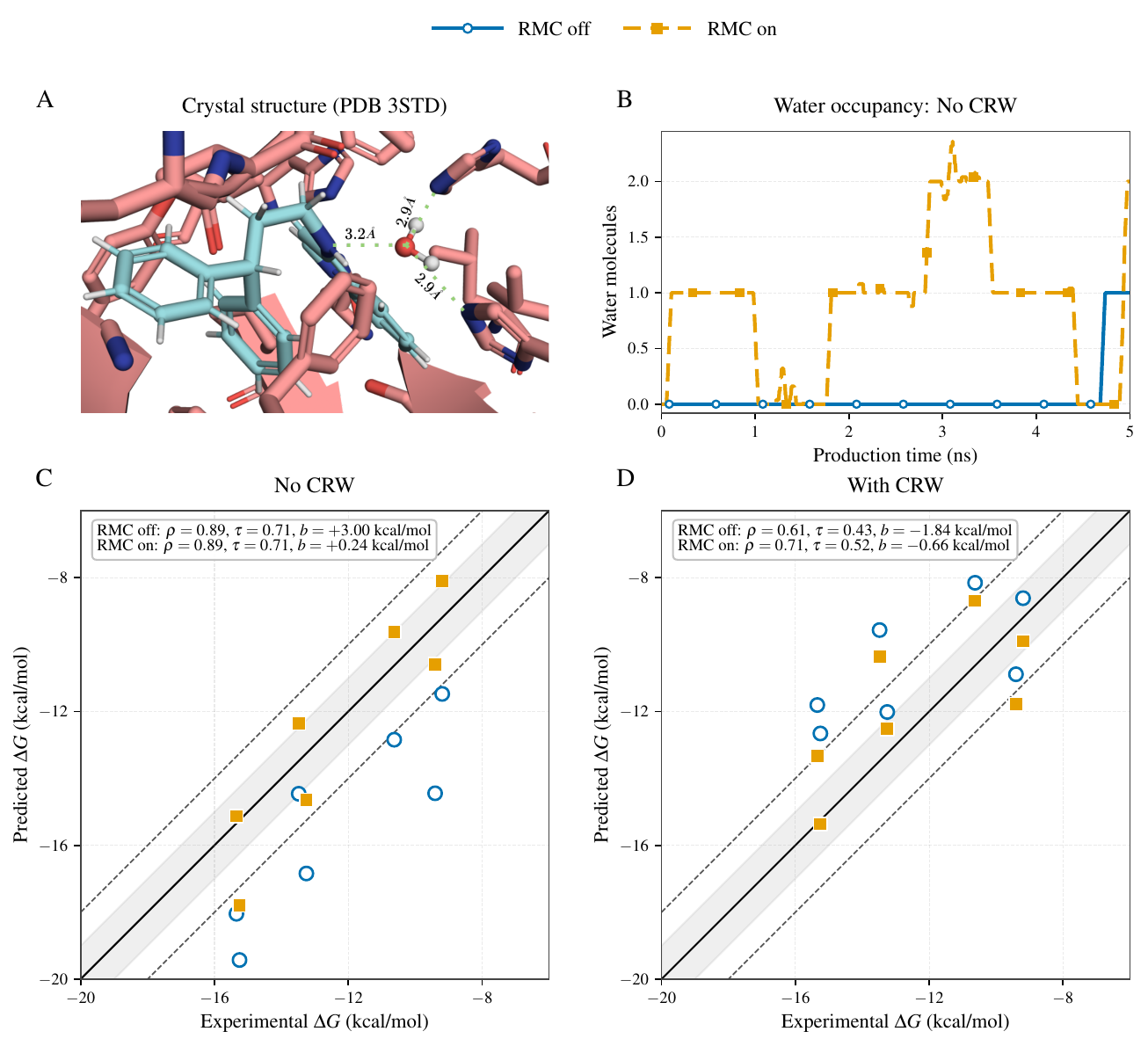}
\caption{
Structure, water occupancy, and binding free-energy predictions for scytalone dehydratase under four water-sampling protocols.
CRW denotes crystallographic waters in the initial structure, and RMC denotes RelocationMC sampling.
(A) Crystal structure of the binding site (PDB ID: 3STD), with the ligand shown in cyan and a bound water molecule shown in red and white.
(B) Water occupancy in simulations initialized without crystallographic waters.
Each curve represents one physical-state trajectory of ligand 3d.
Curves show the centered 25-frame rolling mean of the number of water molecules within \SI{3.5}{\angstrom} of the monitored site during the complete \SI{5}{\nano\second} production trajectory.
The corresponding simulations initialized with crystallographic waters are shown in Fig.~\ref{fig:scyt_dehyd_water_with_crw} in the appendix.
(C, D) Predicted and experimental binding free energies for seven ligands, from simulations initialized without and with crystallographic waters, respectively.
Points show the mean across three independent ABFE calculations.
Blue circles and orange squares indicate simulations with RelocationMC disabled and enabled, respectively.
Insets report Spearman's $\rho$, Kendall's $\tau$, and the signed offset $b$ (in \si{\kcal\per\mole}) for the mean predictions under each protocol.
The solid diagonal indicates perfect agreement, the gray band denotes deviations within \SI{1}{\kcal\per\mole}, and the dashed lines denote deviations of \SI{2}{\kcal\per\mole}.
}
\label{fig:scyt_dehyd_water_case_study}
\end{figure}

The crystal structure contains a buried water molecule that forms three hydrogen bonds: one with a ligand nitrogen and one each with the NE2 atoms of His79 and His104 (Fig.~\ref{fig:scyt_dehyd_water_case_study}A).
We defined the cavity center as the minimum-image geometric center of these three nitrogen atoms.
Water occupancy was calculated as the number of water molecules within \SI{3.5}{\angstrom} of this center during the production trajectory.
We performed four \SI{5}{\nano\second} simulations of ligand 3d that combined the presence or absence of crystallographic waters with RelocationMC enabled or disabled.
Each condition used one fully coupled physical-state NPT trajectory at \SI{298.15}{\kelvin} and \SI{1.01325}{\bar}.
These diagnostic runs used a \SI{2}{\femto\second} integration time step, normal hydrogen masses, and a \SI{2}{\pico\second} output interval.
They are separate from the HREMD ABFE production trajectories.
Condition-specific RelocationMC settings are given in Appendix~\ref{sec:scyt_dehyd_water_with_crw}.

Without crystallographic waters or RelocationMC, the cavity remained essentially dry for most of the trajectory (Fig.~\ref{fig:scyt_dehyd_water_case_study}B).
Enabling RelocationMC hydrated the initially dry cavity within approximately \SI{0.2}{\nano\second} and produced discrete transitions between occupancy states.
The corresponding trajectories initialized with crystallographic waters are shown in Fig.~\ref{fig:scyt_dehyd_water_with_crw} in the appendix.
The observed rapid hydration illustrates the effect of RelocationMC in this diagnostic run, although the protocol-dependent occupancy distributions suggest that \SI{5}{\nano\second} may still be insufficient for complete equilibration.

We next assessed the effect of RelocationMC on ABFE predictions for the seven scytalone dehydratase ligands in the Schr\"odinger benchmark set (Fig.~\ref{fig:scyt_dehyd_water_case_study}C,D).
Each protocol combined the presence or absence of crystallographic waters with RelocationMC enabled or disabled, and the plotted values are averages over three independent calculations.
We quantified the signed series-level offset as $b=\langle\Delta G_{\mathrm{exp}}-\Delta G_{\mathrm{pred}}\rangle$.
Without RelocationMC, this offset was \SI{3.00}{\kcal\per\mole} without crystallographic waters and \SI{-1.84}{\kcal\per\mole} with crystallographic waters.
Enabling RelocationMC reduced the corresponding offsets to \SI{0.24}{\kcal\per\mole} and \SI{-0.66}{\kcal\per\mole}, respectively.
RelocationMC therefore reduced the dependence of the series-level offset on the initial water configuration for this ligand set.
The changes in squared Pearson correlation ($R^2$) were more modest: from 0.77 to 0.84 without crystallographic waters and from 0.49 to 0.51 with crystallographic waters.
Spearman's $\rho$ remained 0.89 without crystallographic waters and increased from 0.61 to 0.71 with crystallographic waters when RelocationMC was enabled.
Because this comparison contains only seven ligands with similar binding geometries, the correlation differences should be interpreted cautiously.

Without crystallographic waters, enabling RelocationMC changed the offset-adjusted MAE from \SI{1.09}{\kcal\per\mole} to \SI{1.26}{\kcal\per\mole} and RMSE from \SI{1.26}{\kcal\per\mole} to \SI{1.36}{\kcal\per\mole}.
With crystallographic waters, the corresponding errors remained approximately \SI{1.48}{\kcal\per\mole} and \SI{1.74}{\kcal\per\mole}, respectively.
RelocationMC therefore reduced the signed offsets for this series without improving the offset-adjusted errors.
The lower correlations for calculations initialized with crystallographic waters indicate that initialization-dependent differences remain, but this comparison does not identify their origin or establish which protocol has equilibrated.

We also compared calculations with and without RelocationMC across all 43 retained Schr\"odinger benchmark systems, keeping all other settings identical (Appendix~\ref{sec:rmc_benchmark_comparison}).
Relative to baseline calculations, RelocationMC reduced the absolute offset by at least \SI{0.1}{\kcal\per\mole} in 19 systems, with slightly higher macro-averaged rank correlations and lower offset-adjusted errors (Fig.~\ref{fig:rmc_system_signed_offset} and Table~\ref{tab:rmc_benchmark_comparison}).
These results indicate that offset reductions extend beyond the scytalone dehydratase case without a loss of aggregate predictive performance, although individual systems respond differently.

\subsection{Benchmark results}

We evaluated Felis ABFE predictions for 857 ligand entries across 43 benchmark systems.
Accounting for alternative binding poses and protonation states present in the original repository, this amounts to 998 protein--ligand configurations in total.
We compared our results against state-of-the-art RBFE results~\cite{rossMaximalCurrentFEPBenchmark2023}, calculated by FEP+ paired with the OPLS4~\cite{luOPLS4ImprovingForce2021} force field.
The referenced FEP+ RBFE results used a \SI{20}{\nano\second} simulation protocol.
We performed three independent ABFE calculations with Felis for each protein--ligand configuration, with \SI{5}{\nano\second} of production sampling per thermodynamic state.
The \SI{3}{\nano\second} estimates were recalculated from the first \SI{3}{\nano\second} of the same trajectories, so the two durations represent paired analyses rather than independent simulation sets.
The production Felis ABFE calculations retained the curated crystallographic waters and enabled RelocationMC sampling in the protein--ligand leg.
Both Felis ABFE and FEP+ RBFE predictions were derived from variant-level calculations and compared against experimental measurements.
In the public FEP+ RBFE dataset, ligand values were obtained by first applying cycle-closure correction (CCC) on the RBFE network.
Next, ligand variants were combined into a single prediction per molecule using conformer and rotamer symmetry handling, pKa/tautomer corrections, and solvent free-energy corrections.
RBFE predictions can be placed on an absolute affinity scale by matching the mean predicted and experimental values within each system.
Felis used the same variant-level postprocessing to obtain ligand-level ABFE predictions, but did not apply CCC.
To quantify performance within each target, we compared Spearman's $\rho$, Kendall's $\tau$, offset-adjusted mean absolute error (MAE), and offset-adjusted root mean square error (RMSE).
For each ligand series, we defined the signed offset as $b=\langle\Delta G_{\mathrm{exp}}-\Delta G_{\mathrm{pred}}\rangle$ and the offset-adjusted prediction as $\Delta G_{\mathrm{pred}}+b$.
The Felis MAE and RMSE are computed from these offset-adjusted ligand-level predictions.
FEP+ system-level metrics are taken from the published benchmark data and used as supplied, without additional offset adjustment in our analysis.
They are not directly comparable to the pairwise or edgewise errors reported for the RBFE benchmark~\cite{rossMaximalCurrentFEPBenchmark2023}.

\begin{figure}[t]
\centering
\includegraphics[width=0.8\linewidth]{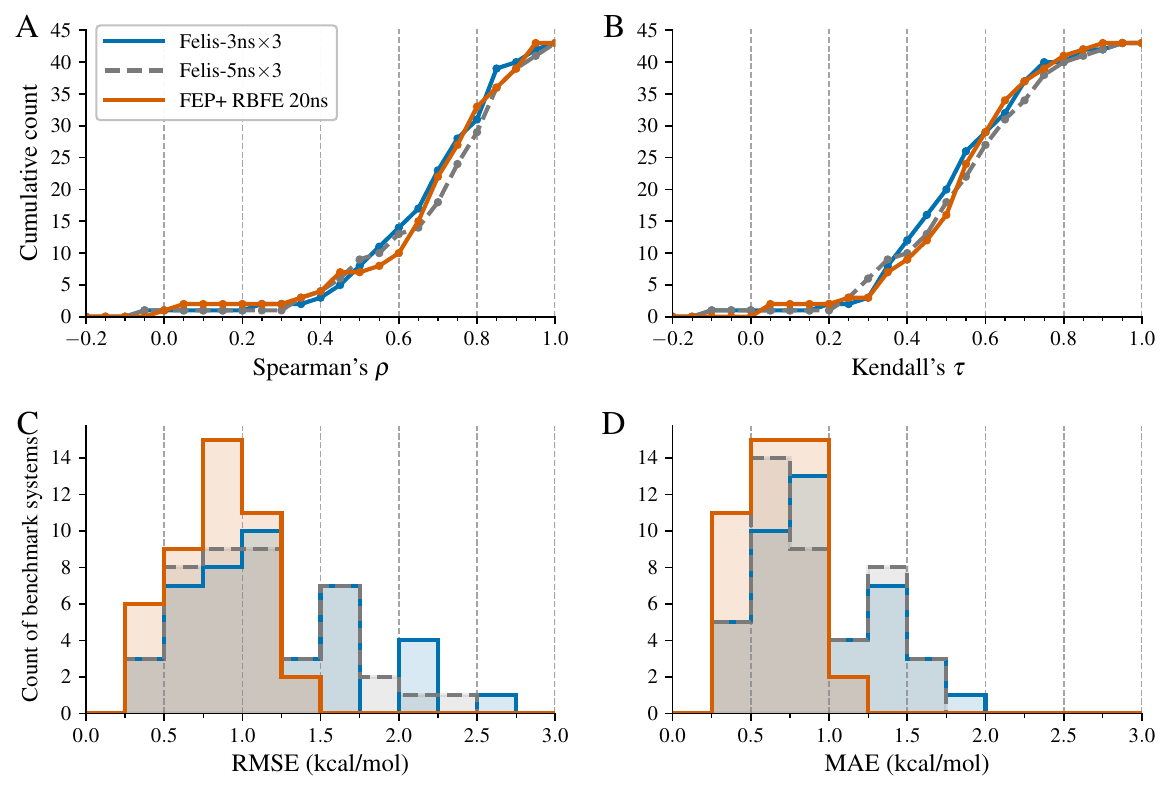}
\caption{
Distributions of within-system rank correlations and offset-adjusted errors for Felis ABFE and FEP+ RBFE across 43 benchmark systems comprising 857 ligand entries.
Panels A--B show cumulative counts of benchmark systems for Spearman's $\rho$ and Kendall's $\tau$, respectively.
Panels C--D show histograms of offset-adjusted RMSE and MAE, respectively.
The Felis ABFE calculations retained curated crystallographic waters and enabled RelocationMC in the protein--ligand leg.
Each prediction averages three independent calculations, with \SI{3}{\nano\second} and \SI{5}{\nano\second} estimates obtained from the first \SI{3}{\nano\second} and the full \SI{5}{\nano\second} of the same production trajectories, respectively.
For Felis, MAE and RMSE are calculated after applying the signed offset within each system.
FEP+ system-level metrics are reproduced from the published benchmark data without further offset adjustment.
The referenced FEP+ RBFE results used \SI{20}{\nano\second} sampling.
}
\label{fig:ranking_performance}
\end{figure}

Spearman's $\rho$ evaluates rank correlation, while Kendall's $\tau$ focuses on concordant and discordant pairs.
Macro-averages weight each benchmark system equally.
We estimated pointwise 95\% confidence intervals (CIs) by paired resampling of target clusters, keeping systems that share a protein target together.
At \SI{5}{\nano\second}, Felis ABFE predictions yielded macro-averaged Spearman's $\rho=0.68$ (95\% CI $[0.62,0.74]$) and Kendall's $\tau=0.53$ (95\% CI $[0.47,0.60]$) across the 43 benchmark systems (Fig.~\ref{fig:ranking_performance}).
The corresponding macro-averaged offset-adjusted MAE was \SI{0.89}{\kcal\per\mole} (95\% CI $[0.80,0.99]$).
The macro-averaged offset-adjusted RMSE was \SI{1.12}{\kcal\per\mole} (95\% CI $[1.00,1.25]$).
The \SI{3}{\nano\second} analysis yielded macro-averaged Spearman's $\rho=0.66$, Kendall's $\tau=0.51$, offset-adjusted MAE of \SI{0.93}{\kcal\per\mole}, and offset-adjusted RMSE of \SI{1.18}{\kcal\per\mole}.
Intervals for the macro-averages and paired differences are reported in Tables~\ref{tab:benchmark_macro_ci} and~\ref{tab:benchmark_paired_ci}, with the procedure described in Appendix~\ref{sec:benchmark_uncertainty}.
Extending the analyzed interval from \SI{3}{\nano\second} to \SI{5}{\nano\second} modestly improved these averages, although this does not establish convergence for every system.
For the same benchmark systems, FEP+ RBFE yielded macro-averaged Spearman's $\rho=0.66$, Kendall's $\tau=0.52$, offset-adjusted MAE of \SI{0.68}{\kcal\per\mole}, and offset-adjusted RMSE of \SI{0.84}{\kcal\per\mole}.
Felis ABFE predictions therefore showed similar average rank correlations to FEP+ RBFE but larger offset-adjusted errors, with substantial variation among systems.
This comparison is challenging for ABFE because the dataset was designed for RBFE benchmarking and the dynamic range of ligand affinities is limited for several targets.
Sampling variability can be comparable to the affinity range within a target, motivating the use of three independent calculations to reduce uncertainty in the mean predictions.

Per-system results are reported in Tables~\ref{tab:schr_abfep_stats_3ns_joint25}, \ref{tab:schr_abfep_stats_5ns_joint25}, and~\ref{tab:fep_plus_rb_prediction_metrics}.
Both methods performed poorly for the hc\_bace\_2 system from the opls\_stress dataset.
At \SI{5}{\nano\second}, Felis ABFE predictions yielded Spearman's $\rho=-0.07$ and Kendall's $\tau=-0.13$, while FEP+ RBFE yielded Spearman's $\rho=-0.04$ and Kendall's $\tau=0.00$.
Except for the strongest and weakest binders, the experimental measurements for this system span only \SI{0.4}{\kcal\per\mole}.
Given typical experimental uncertainty, such a narrow range makes ranking metrics difficult to interpret.

\begin{figure}[t]
\centering
\includegraphics[width=0.8\linewidth]{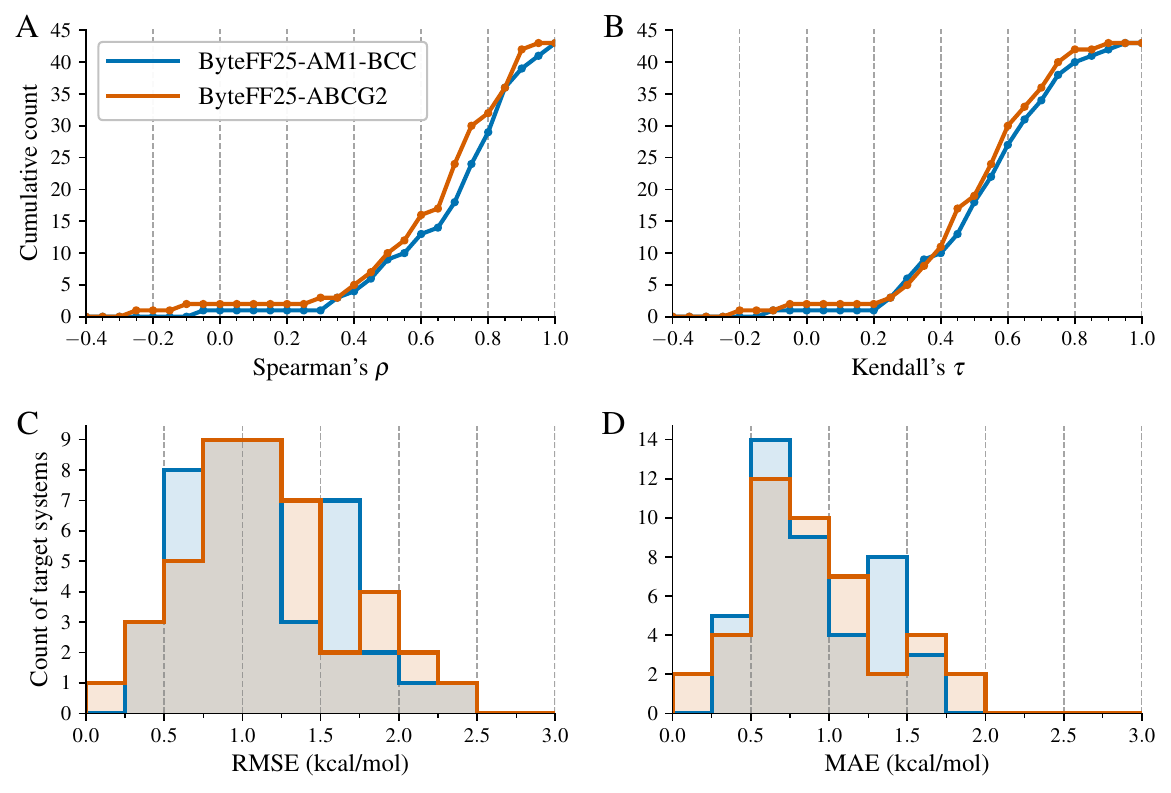}
\caption{
Distributions of within-system rank correlations and offset-adjusted errors for ByteFF25-AM1-BCC and ByteFF25-ABCG2 across 43 benchmark systems comprising 857 ligand entries.
Felis ABFE predictions average three independent calculations with \SI{5}{\nano\second} of production sampling per thermodynamic state.
The two force fields were trained on the same quantum chemistry data using their respective charge schemes.
The comparison includes differences in bonded and van der Waals parameters as well as partial atomic charges.
}
\label{fig:ranking_performance_ligand_ff}
\end{figure}

In addition to the comparison against state-of-the-art RBFE, we compared ByteFF25 parameterizations based on different ligand charge models.
The preceding benchmark calculations used ByteFF25-AM1-BCC to parameterize ligands and cofactors.
AM1-BCC was originally developed as a rapid approximation to the HF/6-31G* RESP charge model and has been widely used for simulations of drug-like molecules~\cite{jakalianFastEfficientGeneration2002}.
Recently, the ABCG2 charge model~\cite{heABCG22023,heABCG22025} has been proposed as an alternative to AM1-BCC, with improved accuracy for experimental solvation free energies of organic molecules.
Using the same training data and protocol, we trained ByteFF25-ABCG2 and compared the two ByteFF25 parameterizations.
The parameterizations differ in bonded and Lennard-Jones parameters as well as partial atomic charges.
We therefore interpret the results as a comparison of complete force-field variants.
At \SI{5}{\nano\second}, ByteFF25-AM1-BCC yielded higher macro-averaged Spearman's $\rho$ and Kendall's $\tau$ (0.68 and 0.53) than ByteFF25-ABCG2 (0.63 and 0.50; Fig.~\ref{fig:ranking_performance_ligand_ff}).
Their offset-adjusted MAE and RMSE were similar: \SI{0.89}{\kcal\per\mole} and \SI{1.12}{\kcal\per\mole} for AM1-BCC, compared with \SI{0.90}{\kcal\per\mole} and \SI{1.14}{\kcal\per\mole} for ABCG2.
The paired 95\% CIs include zero for all four metrics (Table~\ref{tab:benchmark_paired_ci}), so the observed differences provide limited evidence of a performance advantage for either force-field variant.
These intervals do not establish equivalence between the force-field variants.
Complete per-system metrics for ByteFF25-ABCG2 are provided in Table~\ref{tab:schr_abfep_stats_5ns_abcg2}.
A previous study also compared these models on a smaller benchmark~\cite{beheraEvaluationABCG2Charge2025}.
A controlled comparison that holds bonded and Lennard-Jones parameters fixed would be needed to isolate charge-model effects.

\subsection{Case study on a challenging KRAS(G12D) dataset}

Following large-scale benchmarking, we extended our evaluation to a more challenging case study: the KRAS(G12D) ligand dataset.
KRAS is one of the most frequently mutated oncogenes in human cancers and acts as a key molecular switch for cell proliferation and survival.
Designing effective inhibitors for the KRAS(G12D) mutant has historically been challenging, in part because of the shallow binding pocket and strong affinity for native GTP/GDP substrates.
MRTX1133 inhibits phosphorylation of extracellular signal-regulated kinase (ERK) in AGS cells with an IC\textsubscript{50} of \SI{2}{\nano\Molar}~\cite{wangDiscoveryMRTX11332022}.
Its reported surface plasmon resonance (SPR) binding-affinity estimate is $K_D\approx\SI{0.2}{\pico\Molar}$~\cite{wangDiscoveryMRTX11332022,hallin2022anti}, although the primary report notes that this value is beyond the instrument's accurate determination range.
We modeled MRTX1133 in a dicationic protonation state, with its protonated amines positioned to interact with Asp12 and Glu62 in the cocrystal binding mode (PDB ID: 7RPZ).

We selected a congeneric series of ten ligands, including MRTX1133, from the Uni-FEP dataset~\cite{zouBreakingBarriersFEP2025}.
For MRTX1133, we retained the Uni-FEP benchmark reference of \SI{-12.81}{\kcal\per\mole}.
In this subset, ligand SMILES strings, experimental affinities, and the protein structure of the target are provided in the benchmark dataset.
Protein residue protonation states were assigned at $\mathrm{pH}=7.4$ using PROPKA 3 (v3.5)~\cite{olssonPROPKA3ConsistentTreatment2011,sondergaardImprovedTreatmentLigands2011} in the presence of bound GDP, a chelated Mg$^{2+}$ ion, and the Uni-FEP reference ligand.
Ligand poses and protonation states were manually curated to maintain consistency with the MRTX1133 cocrystal binding mode (Fig.~\ref{fig:kras_ligands}).
This series challenges thermodynamic sampling in the shallow, dynamic pocket near switch-II and in the charged environment formed by the cofactor, Mg$^{2+}$ ion, and ligands.
Based on prior force-field comparisons, we used ByteFF25-AM1-BCC for ligands and cofactors together with AMBER ff14SB for the protein and followed the same protocol as the large-scale benchmark.
Curated variants were combined independently within each run using Felis postprocessing, and the resulting ligand-level predictions were averaged across three runs.
Alternative ligand variants are likewise included in the recent ToxBench synthetic dataset~\cite{liuToxBenchBindingAffinity2025}.

\begin{figure}[htbp]
    \centering
    \includegraphics[width=0.8\linewidth]{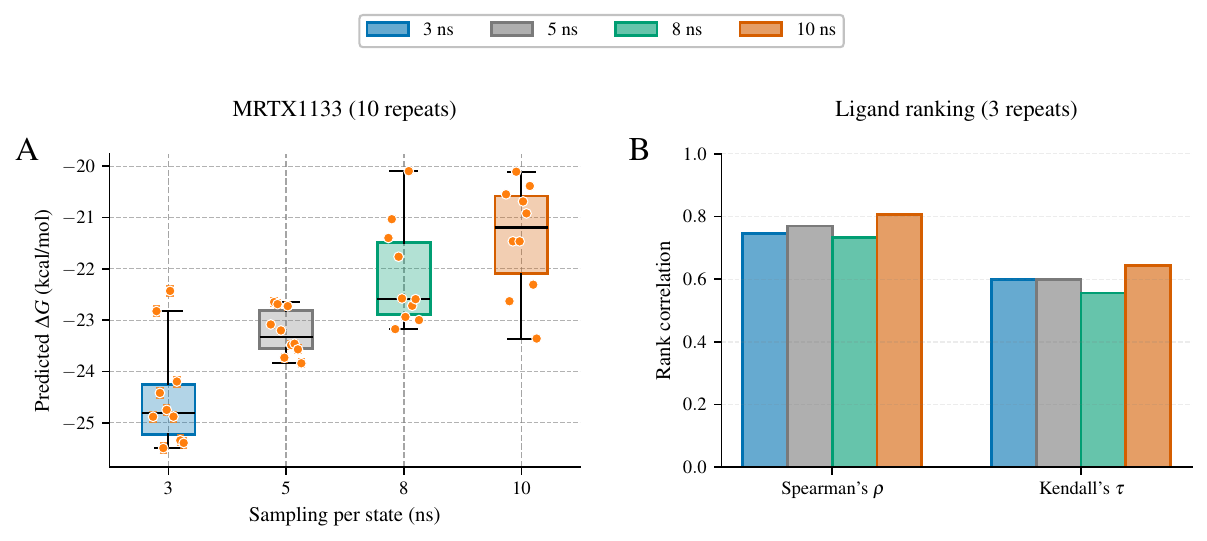}
    \caption{
    Sampling-length dependence and ligand ranking for the KRAS(G12D) series.
    (A) MRTX1133 binding free energy estimates from ten independent ABFE calculations at each displayed sampling duration.
    Boxes show the median and interquartile range, with individual estimates overlaid.
    Vertical bars indicate within-calculation BAR uncertainties combined in quadrature from the two alchemical legs.
    (B) Spearman's $\rho$ and Kendall's $\tau$ for the ten-ligand series at \SIlist{3;5;8;10}{\nano\second} of production sampling per thermodynamic state.
    Ranking metrics use the mean of three postprocessed ligand-level predictions and the Uni-FEP experimental reference values.
    }
    \label{fig:kras}
\end{figure}

We evaluated the ten-ligand series at \SIlist{3;5;8;10}{\nano\second}, using three independent calculations per ligand variant at each duration.
MRTX1133 sampling-length dependence was assessed using ten independent calculations at each of \SIlist{3;5;8;10}{\nano\second}.
The corresponding convergence, neighboring-state overlap, and replica-mixing diagnostics are provided in the accompanying raw data, as described in Appendix~\ref{sec:kras_details}.
Figure~\ref{fig:kras}A displays the \SIlist{3;5;8;10}{\nano\second} estimates.
The mean MRTX1133 predictions were \SIlist{-24.46;-23.24;-22.13;-21.39}{\kcal\per\mole} at \SIlist{3;5;8;10}{\nano\second}, respectively, becoming less favorable with longer sampling.
At \SI{10}{\nano\second}, the across-run standard deviation was \SI{1.07}{\kcal\per\mole}.
The completed \SI{8}{\nano\second} repeat set yielded a mean of \SI{-22.13}{\kcal\per\mole} and an across-run standard deviation of \SI{1.02}{\kcal\per\mole}.
The change in the mean and the remaining variability warrant caution when assessing convergence.

Across all four sampling durations, Spearman's $\rho$ ranged from 0.73 to 0.81 and Kendall's $\tau$ from 0.56 to 0.64.
At \SI{10}{\nano\second}, Felis ABFE predictions yielded Spearman's $\rho=0.81$ and Kendall's $\tau=0.64$.
Offset-adjusted MAE decreased from \SI{1.74}{\kcal\per\mole} at \SI{3}{\nano\second} to \SI{1.47}{\kcal\per\mole} at \SI{10}{\nano\second}, while offset-adjusted RMSE decreased from \SI{2.42}{\kcal\per\mole} to \SI{2.08}{\kcal\per\mole}.
The \SI{8}{\nano\second} results in Fig.~\ref{fig:kras_scatter}E gave similar offset-adjusted errors, with MAE of \SI{1.47}{\kcal\per\mole} and RMSE of \SI{2.02}{\kcal\per\mole}.
Rank correlations varied modestly across these durations, while the mean unadjusted MRTX1133 binding free energy became less favorable with longer sampling.
Ligand-level results for all four durations are shown in Fig.~\ref{fig:kras_scatter}.

Before applying the signed offset, the KRAS(G12D) predictions were systematically more favorable than the experimental references.
The signed offset $b=\langle\Delta G_{\mathrm{exp}}-\Delta G_{\mathrm{pred}}\rangle$ decreased from \SI{6.12}{\kcal\per\mole} at \SI{3}{\nano\second} to \SI{3.71}{\kcal\per\mole} at \SI{10}{\nano\second}.
We applied this offset separately at each duration to assess offset-adjusted ligand-level errors.
Protein reorganization has been proposed as a contributor to such offsets~\cite{chenEnhancingHitDiscovery2023}, but its contribution was not isolated here.
Appendix~\ref{sec:kras_details} discusses the remaining discrepancies and the reported SPR affinity estimate for MRTX1133.

\FloatBarrier

\section{Discussion and conclusion}

In this work, we presented Felis, an automated and scalable source-available toolkit for high-throughput absolute binding free energy (ABFE) predictions.
Across 43 benchmark systems and 857 ligand entries, ABFE calculations with Felis showed average rank correlations similar to those of FEP+ RBFE, although their offset-adjusted errors were larger.
The comparison used three independent \SI{5}{\nano\second} ABFE calculations per configuration, with \SI{3}{\nano\second} estimates obtained by analyzing the initial portion of the same trajectories.
The nominal sampling duration per thermodynamic state was shorter than the referenced \SI{20}{\nano\second} FEP+ RBFE protocol, but these durations alone do not establish a computational cost advantage.
Neither force-field parameters nor alchemical schedules were adjusted using experimental affinities from the benchmarks.
The KRAS(G12D) case study extends this evaluation to highly charged ligands, with rank correlations maintained across sampling durations despite substantial series-level offsets.
Collectively, these findings support Felis as a practical ABFE toolkit for ligand prioritization, while identifying limitations that require further validation.

Despite the progress reported here, the current benchmark dataset, while comparable in size to the recent OpenFE collaborative assessment~\cite{baumannLargescaleCollaborativeAssessment2025}, is not yet sufficient to characterize performance across the full diversity of targets and chemotypes encountered in real-world prospective projects.
A much larger benchmark dataset has been released recently by Uni-FEP~\cite{zouBreakingBarriersFEP2025}.
Extending the Felis evaluation to this dataset requires preparation and validation of binding poses and protonation states.
Such an assessment would help distinguish errors associated with structural preparation, sampling, and force-field parameters.
A comprehensive evaluation of Felis on the full Uni-FEP dataset remains future work.

Real-world adoption of ABFE will likely require further improvements to enhance predictive accuracy~\cite{meyBestPracticesAlchemical2020,hahnBestPracticesConstructing2022}.
Priority areas include (i) more reliable binding poses, (ii) robust protonation-state assignment for both proteins and ligands, and (iii) more accurate and transferable force fields.
As with other free-energy workflows, the accuracy of Felis ABFE predictions depends on the quality of the initial binding poses.
While the benchmark datasets employed in this study used high-confidence structural data, limited pose reliability can attenuate predictive accuracy in de novo applications.
We anticipate that integrating enhanced sampling molecular dynamics~\cite{clarkPredictionProteinLigand2016,xuInducedFitDockingEnables2022} with emerging co-folding strategies~\cite{beumingAreDeepLearning2022,coskunUsingAlphaFoldExperimental2024,thalerBoltzABFEFreeEnergy2025} will further improve binding pose accuracy.
Such advances, together with improved protonation-state assignment and force fields, could broaden the applicability of ABFE calculations with Felis to more challenging prospective settings.

Beyond workflow and sampling improvements, further progress could come from refining nonbonded parameters, which govern protein--ligand interactions in ABFE.
This would complement the improvements in bonded interactions demonstrated in the ByteFF24 benchmarks~\cite{zhengByteFF24}.
Pragmatically, this entails co-optimizing the charge model and van der Waals parameters against hydration free energies and condensed-phase liquid properties, following strategies developed in the OPLS and OpenFF efforts~\cite{harderOPLS3ForceField2016,boothroydImprovingForceField2022,boothroydDevelopmentBenchmarkingOpen2023}.
In parallel, expanding the physical mechanisms represented by the force field (for example, virtual sites to capture sigma-hole anisotropy~\cite{roosOPLS3eExtendingForce2019,luOPLS4ImprovingForce2021} and explicit polarizability~\cite{dammOPLS5AdditionPolarizability2024,ponderCurrentStatusAMOEBA2010} for key functional groups) could improve transferability across challenging chemotypes.
Finally, integrating modern machine-learning potentials as corrective models or hybrid components offers a pathway to handle regimes that exceed the expressiveness of fixed functional-form molecular mechanics while preserving the efficiency required for high-throughput ABFE~\cite{pleFennixbio12025}.

Aside from these major improvements, Felis may benefit from the following additional enhancements.
In ABFE studies, computed $\Delta G$ values frequently exhibit a systematic offset relative to experimental measurements, varying by target protein and potentially by ligand series.
Such an offset typically does not degrade ranking performance unless it varies substantially among ligands.
This global offset is sometimes attributed to protein reorganization free energy in prior work, for example in Schr\"odinger ABFE benchmark reports~\cite{chenEnhancingHitDiscovery2023} and recent large-scale ToxBench analyses~\cite{liuToxBenchBindingAffinity2025}.
Whether the offset is solely due to protein reorganization free energy remains an open question and is beyond the scope of this study.
In future work, we plan to incorporate an explicit estimate of protein reorganization free energy into Felis ABFE calculations to assess whether the global offset is reduced.
Enhanced sampling methods may also improve convergence and reliability, although their benefits depend on the system and protocol~\cite{wangReplicaExchangeSolute2011,wangTerminalFlip2025}.

Looking ahead, runtime and cost can be reduced along complementary axes.
The present calculations use hydrogen mass repartitioning (HMR) with the \texttt{LangevinMiddleIntegrator} and a single \SI{4}{\femto\second} integration time step.
Multiple-time stepping (MTS), such as rRESPA~\cite{respa1992}, could provide additional speedups in \texttt{OpenMM} by evaluating fast and slow forces at different frequencies.
Its effects on free-energy accuracy and numerical stability would require validation.
Further speedups are possible by carefully tuning particle-mesh Ewald (PME) convergence parameters, including real-space cutoff, grid spacing, and tolerance, with validation to ensure negligible impact on accuracy.
Alternative approaches to long-range interactions, such as midtown-splines~\cite{predescuMidtownSpline2020}, prolate spheroidal wave functions (PSWFs)~\cite{liangProlateSpheroidalWave2026}, and u-series~\cite{predescuUSeries2020}, may offer further computational improvements.
Their integration into the Felis workflow would require implementation and validation in \texttt{OpenMM}.
On the sampling method side, the Convergence-Adaptive Roundtrip (CAR) method~\cite{yaoCAR2024} may be a promising pathway to allocate effort dynamically along the alchemical transformation path based on convergence signals to further reduce the total sampling time.
Beyond algorithmic improvements, orchestration and parallelization strategies have been shown to increase alchemical throughput at scale, and similar scheduling can be leveraged in Felis~\cite{kutznerGROMACSCloudGlobal2022,gapsysPreExascaleComputingProtein2022}.
Evaluating these options in Felis with conservative safeguards to preserve accuracy is a priority for future work.

\section{Software availability}

The Felis source code is available at \url{https://github.com/ByteDance-Seed/felis}.
The same repository includes ByteMol and the source code and installation instructions for \texttt{PerturbNonbondedPlugin} and \texttt{RelocationMCPlugin}.
Example workflows and benchmark inputs, including precomputed ligand and cofactor parameters, are also provided there.

\section{Acknowledgments}

We thank Chang Han, Xingyuan Xu, and Tianze Zheng for helpful discussions and assistance with training the ByteFF25 force field.

\clearpage

\bibliographystyle{unsrtnat}
\bibliography{main}

\clearpage

\beginappendix

\appendix
\section{Benchmarks}
\subsection{All results for the selected benchmark dataset}

The Schr\"odinger benchmark dataset was originally curated for relative binding free energy (RBFE) calculations.
To compare Felis ABFE and FEP+ RBFE predictions consistently, we followed the data splitting and ligand assignment protocols of the original benchmark.
For example, ligands for the P2Y1 target (GPCRs dataset) were divided into two subsets based on the RBFE mapping algorithm, as specified in the public repository\footnote{\url{https://github.com/schrodinger/public_binding_free_energy_benchmark}}.
We adopted this splitting strategy to evaluate the ranking performance of both methods on identical subsets.
A similar adherence to the original protocol was applied to the OX2 receptor (ox2\_hip\_custcore).
In principle, multiple states of the same ligand (e.g., protonation states or pose variations) should be combined into a single entry with appropriate conformer and rotamer symmetry handling and pKa/tautomer corrections.
For instance, we identified set\_2\_8 and set\_1\_27 as pose variations of the same ligand associated with a single experimental affinity.
However, to preserve consistency with the benchmark dataset, we treated them as distinct entries, following the original assignment.

The following tables report per-system metrics for the complete benchmark of 43 systems and 857 ligand entries.
The \SI{3}{\nano\second} and \SI{5}{\nano\second} AM1-BCC results use the same ligand set and average three independent calculations.
The \SI{3}{\nano\second} estimates were recalculated from the initial \SI{3}{\nano\second} of the \SI{5}{\nano\second} production trajectories.
In addition to the metrics defined in the main text, we report $R^2$, defined as the square of Pearson's correlation coefficient, and the signed offset $b=\langle\Delta G_{\mathrm{exp}}-\Delta G_{\mathrm{pred}}\rangle$.
The Felis offset-adjusted MAE and RMSE are calculated from $\Delta G_{\mathrm{pred}}+b$.
The ``Correct Order Ratio'' is the fraction of correctly ranked ligand pairs among those whose experimental affinities differ by more than \SI{2}{\kcal\per\mole}.
It is reported on a scale from 0 to 1, with N/A indicating systems that contain no eligible pairs.

\begin{table}[htb]
    \centering
    \caption{Felis ABFE statistics for the initial \SI{3}{\nano\second} of three independent production calculations with crystallographic waters and RelocationMC (ByteFF25-AM1-BCC).}
    \label{tab:schr_abfep_stats_3ns_joint25}
    \begin{tabular}{llrrrrrrr}
      \toprule
      Subset        & System                     & $R^2$ & $\rho$ & $\tau$ & Offset & RMSE & MAE  & \makecell{Correct\\Order\\Ratio} \\
      \midrule
      fragments & frag\_liga\_auto & 0.93 & 0.91 & 0.75 & 0.99 & 1.64 & 1.34 & 1.00 \\
      fragments & frag\_mcl1\_noweak & 0.26 & 0.55 & 0.36 & 0.59 & 1.22 & 0.98 & N/A \\
      fragments & frag\_mup1 & 0.82 & 0.85 & 0.68 & -1.09 & 0.86 & 0.70 & 1.00 \\
      fragments & hsp90\_frag\_1ring & 0.68 & 0.85 & 0.68 & 2.56 & 0.42 & 0.34 & N/A \\
      fragments & hsp90\_frag\_2rings & 0.41 & 0.83 & 0.73 & 7.09 & 0.88 & 0.76 & 1.00 \\
      fragments & t4lysozyme\_uvt & 0.58 & 0.70 & 0.55 & -0.18 & 0.69 & 0.49 & 1.00 \\
      gpcrs & a2a\_hip278 & 0.53 & 0.75 & 0.57 & 2.98 & 1.20 & 0.92 & 1.00 \\
      gpcrs & ox2\_hip\_custcore & 0.41 & 0.66 & 0.46 & 2.70 & 1.05 & 0.83 & 0.95 \\
      gpcrs & p2y1\_meta & 0.36 & 0.44 & 0.31 & -5.26 & 0.68 & 0.60 & 0.92 \\
      gpcrs & p2y1\_ortho & 0.78 & 0.81 & 0.63 & -4.27 & 0.47 & 0.36 & 1.00 \\
      jacs & bace & 0.47 & 0.74 & 0.55 & 2.61 & 1.25 & 1.00 & 0.98 \\
      jacs & cdk2 & 0.19 & 0.37 & 0.28 & 2.22 & 1.12 & 0.88 & 0.77 \\
      jacs & jnk1 & 0.42 & 0.69 & 0.48 & 0.56 & 0.80 & 0.65 & 1.00 \\
      jacs & mcl1 & 0.45 & 0.70 & 0.51 & 0.51 & 1.41 & 1.10 & 0.96 \\
      jacs & p38 & 0.47 & 0.74 & 0.58 & 4.46 & 1.24 & 0.84 & 0.96 \\
      jacs & ptp1b & 0.29 & 0.73 & 0.65 & 13.22 & 2.58 & 1.81 & 0.83 \\
      jacs & thrombin & 0.72 & 0.79 & 0.67 & -1.18 & 0.47 & 0.35 & N/A \\
      jacs & tyk2 & 0.64 & 0.83 & 0.67 & 0.23 & 0.89 & 0.64 & 0.97 \\
      janssen\_bace & bace\_ciordia\_prospective & 0.52 & 0.65 & 0.50 & -0.21 & 0.77 & 0.61 & 1.00 \\
      janssen\_bace & bace\_ciordia\_retro & 0.44 & 0.67 & 0.48 & 4.60 & 0.96 & 0.76 & 0.94 \\
      janssen\_bace & bace\_p3\_arg368\_in & 0.30 & 0.55 & 0.33 & 1.78 & 1.57 & 1.24 & 0.83 \\
      macrocycles & 3RKZ\_lig62to70\_alpha05 & 0.77 & 0.89 & 0.81 & 9.10 & 1.72 & 1.40 & 0.92 \\
      macrocycles & hsp90\_3hvd\_custcore & 0.88 & 0.94 & 0.82 & 2.88 & 0.65 & 0.56 & 1.00 \\
      merck & cdk8 & 0.74 & 0.81 & 0.65 & 3.86 & 1.02 & 0.86 & 0.98 \\
      merck & cmet & 0.72 & 0.83 & 0.72 & 7.20 & 1.69 & 1.34 & 0.94 \\
      merck & eg5 & 0.36 & 0.47 & 0.35 & 0.92 & 2.01 & 1.62 & 0.94 \\
      merck & hif2a & 0.13 & 0.48 & 0.34 & 0.58 & 1.68 & 1.26 & 0.73 \\
      merck & pfkfb3 & 0.29 & 0.53 & 0.37 & -0.05 & 1.16 & 0.91 & 0.84 \\
      merck & shp2 & 0.29 & 0.63 & 0.43 & 3.34 & 1.18 & 0.89 & 0.88 \\
      merck & syk & 0.22 & 0.43 & 0.31 & -0.84 & 1.55 & 1.29 & 0.92 \\
      merck & tnks2 & 0.09 & 0.24 & 0.18 & 4.37 & 1.29 & 1.06 & 0.75 \\
      misc & cdk8\_koehler & 0.37 & 0.52 & 0.38 & 1.13 & 0.66 & 0.60 & N/A \\
      misc & galectin3\_extra & 0.62 & 0.74 & 0.55 & 2.69 & 0.91 & 0.74 & 1.00 \\
      misc & hfaah & 0.64 & 0.57 & 0.40 & 1.95 & 1.05 & 0.85 & 0.96 \\
      opls\_stress & fxa\_set\_4 & 0.55 & 0.67 & 0.49 & 1.63 & 0.71 & 0.59 & 1.00 \\
      opls\_stress & fxa\_yoshikawa\_set & 0.57 & 0.64 & 0.50 & 1.77 & 1.73 & 1.37 & 0.96 \\
      opls\_stress & hc\_bace\_1 & 0.60 & 0.96 & 0.90 & 12.51 & 0.65 & 0.60 & 1.00 \\
      opls\_stress & hc\_bace\_2 & 0.17 & -0.07 & -0.13 & 13.98 & 2.21 & 1.45 & 0.00 \\
      waterset & brd41\_ASH106 & 0.26 & 0.45 & 0.36 & 0.61 & 0.94 & 0.82 & 0.67 \\
      waterset & chk1 & 0.78 & 0.82 & 0.66 & 4.81 & 1.06 & 0.77 & 1.00 \\
      waterset & hsp90\_kung & 0.45 & 0.53 & 0.41 & 5.24 & 2.01 & 1.51 & 1.00 \\
      waterset & scyt\_dehyd & 0.36 & 0.64 & 0.43 & -0.26 & 2.06 & 1.70 & 0.85 \\
      waterset & taf12 & 0.50 & 0.76 & 0.57 & 0.62 & 0.56 & 0.49 & 1.00 \\
      \bottomrule
    \end{tabular}
    \begin{flushleft}
    Here, $\rho$ denotes Spearman's $\rho$ and $\tau$ denotes Kendall's $\tau$; offset, offset-adjusted MAE, and offset-adjusted RMSE are reported in \SI{}{\kcal\per\mole}.
    \end{flushleft}
  \end{table}
  \begin{table}[htb]
    \centering
    \caption{Felis ABFE statistics for three independent \SI{5}{\nano\second} production calculations with crystallographic waters and RelocationMC (ByteFF25-AM1-BCC).}
    \label{tab:schr_abfep_stats_5ns_joint25}
    \begin{tabular}{llrrrrrrr}
      \toprule
      Subset        & System                     & $R^2$ & $\rho$ & $\tau$ & Offset & RMSE & MAE  & \makecell{Correct\\Order\\Ratio} \\
      \midrule
      fragments & frag\_liga\_auto & 0.91 & 0.88 & 0.75 & 0.80 & 1.68 & 1.34 & 1.00 \\
      fragments & frag\_mcl1\_noweak & 0.29 & 0.59 & 0.42 & 0.60 & 1.22 & 1.00 & N/A \\
      fragments & frag\_mup1 & 0.89 & 0.88 & 0.78 & -0.86 & 0.73 & 0.63 & 1.00 \\
      fragments & hsp90\_frag\_1ring & 0.68 & 0.90 & 0.78 & 2.40 & 0.45 & 0.39 & N/A \\
      fragments & hsp90\_frag\_2rings & 0.46 & 0.83 & 0.73 & 6.70 & 0.83 & 0.73 & 1.00 \\
      fragments & t4lysozyme\_uvt & 0.64 & 0.76 & 0.61 & -0.78 & 0.62 & 0.46 & 1.00 \\
      gpcrs & a2a\_hip278 & 0.55 & 0.81 & 0.65 & 2.74 & 1.16 & 0.88 & 1.00 \\
      gpcrs & ox2\_hip\_custcore & 0.41 & 0.68 & 0.48 & 1.62 & 1.04 & 0.78 & 0.94 \\
      gpcrs & p2y1\_meta & 0.42 & 0.46 & 0.35 & -5.25 & 0.63 & 0.53 & 1.00 \\
      gpcrs & p2y1\_ortho & 0.78 & 0.72 & 0.56 & -4.69 & 0.48 & 0.40 & 1.00 \\
      jacs & bace & 0.46 & 0.70 & 0.51 & 2.04 & 1.08 & 0.87 & 0.98 \\
      jacs & cdk2 & 0.27 & 0.34 & 0.23 & 1.61 & 1.03 & 0.86 & 0.77 \\
      jacs & jnk1 & 0.43 & 0.74 & 0.52 & 0.29 & 0.83 & 0.63 & 1.00 \\
      jacs & mcl1 & 0.42 & 0.66 & 0.47 & 0.17 & 1.39 & 1.08 & 0.94 \\
      jacs & p38 & 0.50 & 0.74 & 0.59 & 4.05 & 1.12 & 0.74 & 0.96 \\
      jacs & ptp1b & 0.30 & 0.74 & 0.66 & 11.13 & 2.44 & 1.73 & 0.83 \\
      jacs & thrombin & 0.76 & 0.87 & 0.71 & -0.83 & 0.47 & 0.39 & N/A \\
      jacs & tyk2 & 0.69 & 0.82 & 0.68 & 0.04 & 0.82 & 0.56 & 0.97 \\
      janssen\_bace & bace\_ciordia\_prospective & 0.58 & 0.75 & 0.56 & -0.50 & 0.65 & 0.53 & 1.00 \\
      janssen\_bace & bace\_ciordia\_retro & 0.43 & 0.66 & 0.48 & 3.94 & 0.95 & 0.76 & 0.94 \\
      janssen\_bace & bace\_p3\_arg368\_in & 0.32 & 0.52 & 0.32 & 1.17 & 1.45 & 1.11 & 0.81 \\
      macrocycles & 3RKZ\_lig62to70\_alpha05 & 0.82 & 0.96 & 0.90 & 9.09 & 1.72 & 1.53 & 1.00 \\
      macrocycles & hsp90\_3hvd\_custcore & 0.88 & 0.95 & 0.87 & 2.86 & 0.63 & 0.55 & 1.00 \\
      merck & cdk8 & 0.72 & 0.76 & 0.59 & 3.48 & 1.00 & 0.83 & 0.98 \\
      merck & cmet & 0.73 & 0.83 & 0.72 & 6.59 & 1.61 & 1.28 & 0.94 \\
      merck & eg5 & 0.37 & 0.50 & 0.38 & 0.71 & 1.90 & 1.51 & 0.94 \\
      merck & hif2a & 0.13 & 0.42 & 0.29 & 0.72 & 1.66 & 1.30 & 0.68 \\
      merck & pfkfb3 & 0.31 & 0.49 & 0.35 & -0.73 & 1.08 & 0.84 & 0.85 \\
      merck & shp2 & 0.35 & 0.65 & 0.44 & 3.23 & 1.11 & 0.87 & 0.90 \\
      merck & syk & 0.21 & 0.43 & 0.30 & -0.59 & 1.52 & 1.27 & 0.90 \\
      merck & tnks2 & 0.12 & 0.34 & 0.24 & 3.46 & 1.26 & 1.02 & 0.84 \\
      misc & cdk8\_koehler & 0.29 & 0.36 & 0.29 & 0.66 & 0.82 & 0.72 & N/A \\
      misc & galectin3\_extra & 0.63 & 0.75 & 0.55 & 2.92 & 0.87 & 0.69 & 1.00 \\
      misc & hfaah & 0.62 & 0.58 & 0.43 & 1.25 & 1.07 & 0.88 & 0.95 \\
      opls\_stress & fxa\_set\_4 & 0.65 & 0.76 & 0.53 & 1.34 & 0.63 & 0.51 & 1.00 \\
      opls\_stress & fxa\_yoshikawa\_set & 0.54 & 0.62 & 0.47 & 1.79 & 1.79 & 1.42 & 0.95 \\
      opls\_stress & hc\_bace\_1 & 0.66 & 0.93 & 0.81 & 12.00 & 0.59 & 0.54 & 1.00 \\
      opls\_stress & hc\_bace\_2 & 0.17 & -0.07 & -0.13 & 13.53 & 2.10 & 1.40 & 0.00 \\
      waterset & brd41\_ASH106 & 0.38 & 0.81 & 0.64 & 0.73 & 0.81 & 0.74 & 1.00 \\
      waterset & chk1 & 0.78 & 0.83 & 0.66 & 4.33 & 0.98 & 0.65 & 1.00 \\
      waterset & hsp90\_kung & 0.55 & 0.57 & 0.49 & 4.62 & 1.69 & 1.32 & 1.00 \\
      waterset & scyt\_dehyd & 0.51 & 0.71 & 0.52 & -0.66 & 1.74 & 1.48 & 0.92 \\
      waterset & taf12 & 0.46 & 0.83 & 0.64 & 0.24 & 0.57 & 0.47 & 1.00 \\
      \bottomrule
    \end{tabular}
    \begin{flushleft}
    Here, $\rho$ denotes Spearman's $\rho$ and $\tau$ denotes Kendall's $\tau$; offset, offset-adjusted MAE, and offset-adjusted RMSE are reported in \SI{}{\kcal\per\mole}.
    \end{flushleft}
  \end{table}
  \begin{table}[htb]
    \centering
    \caption{Felis ABFE statistics for three independent \SI{5}{\nano\second} production calculations with crystallographic waters and RelocationMC (ByteFF25-ABCG2).}
    \label{tab:schr_abfep_stats_5ns_abcg2}
    \begin{tabular}{llrrrrrrr}
      \toprule
      Subset        & System                     & $R^2$ & $\rho$ & $\tau$ & Offset & RMSE & MAE  & \makecell{Correct\\Order\\Ratio} \\
      \midrule
      fragments & frag\_liga\_auto & 0.92 & 0.86 & 0.71 & 1.32 & 1.98 & 1.65 & 1.00 \\
      fragments & frag\_mcl1\_noweak & 0.32 & 0.46 & 0.36 & 0.77 & 1.32 & 1.14 & N/A \\
      fragments & frag\_mup1 & 0.91 & 0.94 & 0.88 & -0.94 & 0.60 & 0.52 & 1.00 \\
      fragments & hsp90\_frag\_1ring & 0.86 & 0.88 & 0.78 & -0.68 & 0.21 & 0.19 & N/A \\
      fragments & hsp90\_frag\_2rings & 0.55 & 0.66 & 0.60 & 3.88 & 1.02 & 0.72 & 1.00 \\
      fragments & t4lysozyme\_uvt & 0.75 & 0.85 & 0.70 & -0.07 & 0.54 & 0.37 & 1.00 \\
      gpcrs & a2a\_hip278 & 0.57 & 0.87 & 0.72 & 2.04 & 1.07 & 0.75 & 0.94 \\
      gpcrs & ox2\_hip\_custcore & 0.43 & 0.62 & 0.46 & 2.22 & 1.10 & 0.86 & 0.92 \\
      gpcrs & p2y1\_meta & 0.17 & 0.29 & 0.23 & -6.41 & 0.79 & 0.71 & 0.85 \\
      gpcrs & p2y1\_ortho & 0.21 & 0.49 & 0.44 & -5.97 & 1.13 & 0.69 & 1.00 \\
      jacs & bace & 0.34 & 0.59 & 0.42 & 4.53 & 1.25 & 1.05 & 0.89 \\
      jacs & cdk2 & 0.40 & 0.55 & 0.37 & -0.43 & 0.91 & 0.75 & 0.90 \\
      jacs & jnk1 & 0.47 & 0.72 & 0.54 & -0.73 & 0.81 & 0.65 & 0.95 \\
      jacs & mcl1 & 0.54 & 0.75 & 0.56 & 0.66 & 1.36 & 1.02 & 0.99 \\
      jacs & p38 & 0.42 & 0.72 & 0.54 & 3.20 & 0.96 & 0.66 & 0.96 \\
      jacs & ptp1b & 0.63 & 0.81 & 0.64 & 10.72 & 1.51 & 1.30 & 0.96 \\
      jacs & thrombin & 0.66 & 0.67 & 0.56 & 0.29 & 0.31 & 0.24 & N/A \\
      jacs & tyk2 & 0.39 & 0.66 & 0.48 & 0.09 & 1.06 & 0.76 & 0.91 \\
      janssen\_bace & bace\_ciordia\_prospective & 0.26 & 0.35 & 0.28 & 1.16 & 0.78 & 0.48 & 0.67 \\
      janssen\_bace & bace\_ciordia\_retro & 0.64 & 0.75 & 0.58 & 5.23 & 0.74 & 0.59 & 0.98 \\
      janssen\_bace & bace\_p3\_arg368\_in & 0.20 & 0.40 & 0.30 & 2.33 & 1.38 & 1.01 & 0.81 \\
      macrocycles & 3RKZ\_lig62to70\_alpha05 & 0.77 & 0.86 & 0.71 & 9.75 & 2.24 & 1.68 & 0.92 \\
      macrocycles & hsp90\_3hvd\_custcore & 0.87 & 0.87 & 0.78 & 1.07 & 0.75 & 0.58 & 1.00 \\
      merck & cdk8 & 0.51 & 0.74 & 0.58 & 3.69 & 1.27 & 0.86 & 0.95 \\
      merck & cmet & 0.70 & 0.83 & 0.67 & 5.91 & 1.16 & 0.90 & 0.96 \\
      merck & eg5 & 0.46 & 0.57 & 0.41 & 0.29 & 1.86 & 1.53 & 0.94 \\
      merck & hif2a & 0.10 & 0.44 & 0.30 & 1.04 & 1.71 & 1.20 & 0.71 \\
      merck & pfkfb3 & 0.34 & 0.57 & 0.41 & -0.55 & 1.16 & 0.90 & 0.89 \\
      merck & shp2 & 0.01 & -0.11 & -0.06 & 2.63 & 2.19 & 1.92 & 0.47 \\
      merck & syk & 0.21 & 0.49 & 0.34 & -3.15 & 1.10 & 0.88 & 0.91 \\
      merck & tnks2 & 0.19 & 0.38 & 0.27 & 2.97 & 1.29 & 1.05 & 0.82 \\
      misc & cdk8\_koehler & 0.56 & 0.73 & 0.56 & 0.58 & 0.70 & 0.62 & N/A \\
      misc & galectin3\_extra & 0.61 & 0.77 & 0.63 & 2.09 & 0.97 & 0.72 & 1.00 \\
      misc & hfaah & 0.73 & 0.69 & 0.53 & 4.46 & 0.77 & 0.59 & 0.95 \\
      opls\_stress & fxa\_set\_4 & 0.63 & 0.73 & 0.53 & 0.22 & 0.65 & 0.58 & 1.00 \\
      opls\_stress & fxa\_yoshikawa\_set & 0.39 & 0.55 & 0.40 & 1.19 & 1.94 & 1.58 & 0.89 \\
      opls\_stress & hc\_bace\_1 & 0.85 & 0.68 & 0.52 & 10.44 & 0.40 & 0.31 & 1.00 \\
      opls\_stress & hc\_bace\_2 & 0.18 & -0.26 & -0.22 & 11.79 & 1.91 & 1.44 & 0.00 \\
      waterset & brd41\_ASH106 & 0.27 & 0.67 & 0.43 & -0.65 & 0.84 & 0.78 & 1.00 \\
      waterset & chk1 & 0.74 & 0.84 & 0.68 & 4.39 & 1.12 & 0.76 & 1.00 \\
      waterset & hsp90\_kung & 0.65 & 0.70 & 0.60 & 3.63 & 2.35 & 1.96 & 1.00 \\
      waterset & scyt\_dehyd & 0.76 & 0.89 & 0.71 & -0.03 & 1.31 & 1.16 & 1.00 \\
      waterset & taf12 & 0.63 & 0.60 & 0.43 & -1.30 & 0.48 & 0.39 & 1.00 \\
      \bottomrule
    \end{tabular}
    \begin{flushleft}
    Here, $\rho$ denotes Spearman's $\rho$ and $\tau$ denotes Kendall's $\tau$; offset, offset-adjusted MAE, and offset-adjusted RMSE are reported in \SI{}{\kcal\per\mole}.
    \end{flushleft}
  \end{table}
  \begin{table}[htb]
    \centering
    \caption{FEP+ \SI{20}{\nano\second} RBFE prediction metrics}
    \label{tab:fep_plus_rb_prediction_metrics}
    \begin{tabular}{llrrrrrrrr}
      \toprule
      Subset        & System                     & \# & $R^2$   & $\rho$ & $\tau$ & Offset & RMSE & MAE  & \makecell{Correct\\Order\\Ratio}\\
      \midrule
      fragments     & frag\_liga\_auto           & 11 & 0.93  & 0.92   & 0.78   & 0.00  & 0.62 & 0.46 & 1.00                             \\
      fragments     & frag\_mcl1\_noweak         & 12 & 0.31  & 0.41   & 0.30   & 0.00  & 0.87 & 0.69 & N/A \\
      fragments     & frag\_mup1                 & 7  & 0.92  & 0.94   & 0.88   & 0.00  & 0.35 & 0.31 & 1.00                             \\
      fragments     & hsp90\_frag\_1ring         & 7  & 0.53  & 0.74   & 0.59   & 0.00  & 0.43 & 0.37 & N/A \\
      fragments     & hsp90\_frag\_2rings        & 6  & 0.25  & 0.31   & 0.20   & 0.00  & 1.02 & 0.91 & 1.00                             \\
      fragments     & t4lysozyme\_uvt            & 12 & 0.43  & 0.55   & 0.36   & 0.00  & 0.61 & 0.51 & 1.00                             \\
      gpcrs         & a2a\_hip278                & 17 & 0.50  & 0.73   & 0.59   & 0.16  & 1.08 & 0.79 & 1.00                             \\
      gpcrs         & ox2\_hip\_custcore         & 51 & 0.33  & 0.62   & 0.44   & 0.00  & 1.13 & 0.91 & 0.83                             \\
      gpcrs         & p2y1\_meta                 & 20 & 0.33  & 0.42   & 0.31   & -0.04 & 1.04 & 0.92 & 0.92                             \\
      gpcrs         & p2y1\_ortho                & 12 & 0.81  & 0.83   & 0.69   & 0.10  & 0.85 & 0.77 & 1.00                             \\
      jacs          & bace                       & 36 & 0.39  & 0.56   & 0.42   & 0.00  & 0.75 & 0.58 & 1.00                             \\
      jacs          & cdk2                       & 16 & 0.42  & 0.62   & 0.44   & 0.00  & 0.90 & 0.72 & 0.83                             \\
      jacs          & jnk1                       & 21 & 0.71  & 0.88   & 0.71   & 0.00  & 0.57 & 0.46 & 1.00                             \\
      jacs          & mcl1                       & 42 & 0.49  & 0.67   & 0.52   & 0.00  & 0.86 & 0.69 & 0.95                             \\
      jacs          & p38                        & 34 & 0.49  & 0.72   & 0.54   & 0.00  & 0.81 & 0.63 & 0.94                             \\
      jacs          & ptp1b                      & 23 & 0.85  & 0.88   & 0.67   & 0.00  & 0.51 & 0.41 & 1.00                             \\
      jacs          & thrombin                   & 11 & 0.68  & 0.78   & 0.64   & 0.00  & 0.53 & 0.42 & N/A \\
      jacs          & tyk2                       & 16 & 0.86  & 0.93   & 0.81   & 0.00  & 0.47 & 0.37 & 1.00                             \\
      janssen\_bace & bace\_ciordia\_prospective & 9  & 0.59  & 0.82   & 0.61   & 0.00  & 0.80 & 0.50 & 1.00                             \\
      janssen\_bace & bace\_ciordia\_retro       & 32 & 0.56  & 0.70   & 0.52   & -0.04 & 0.80 & 0.63 & 0.94                             \\
      janssen\_bace & bace\_p3\_arg368\_in       & 21 & 0.63  & 0.77   & 0.56   & 0.00  & 1.07 & 0.86 & 0.96                             \\
      macrocycles   & 3RKZ\_lig62to70\_alpha05   & 7  & 0.97  & 0.89   & 0.71   & 0.00  & 0.46 & 0.36 & 1.00                             \\
      macrocycles   & hsp90\_3hvd\_custcore      & 10 & 0.46  & 0.35   & 0.31   & 0.00  & 1.34 & 1.18 & 0.73                             \\
      merck         & cdk8                       & 32 & 0.56  & 0.76   & 0.54   & 0.00  & 1.13 & 0.93 & 0.95                             \\
      merck         & cmet                       & 24 & 0.82  & 0.92   & 0.79   & 0.00  & 0.74 & 0.63 & 0.99                             \\
      merck         & eg5                        & 28 & 0.35  & 0.64   & 0.48   & 0.09  & 0.87 & 0.66 & 0.96                             \\
      merck         & hif2a                      & 41 & 0.61  & 0.65   & 0.48   & 0.00  & 0.77 & 0.66 & 0.97                             \\
      merck         & pfkfb3                     & 40 & 0.58  & 0.76   & 0.55   & 0.00  & 0.99 & 0.77 & 0.98                             \\
      merck         & shp2                       & 26 & 0.50  & 0.74   & 0.56   & 0.00  & 1.11 & 0.92 & 0.94                             \\
      merck         & syk                        & 44 & 0.50  & 0.65   & 0.46   & -0.04 & 0.73 & 0.58 & 0.97                             \\
      merck         & tnks2                      & 27 & 0.29  & 0.45   & 0.33   & 0.29  & 1.11 & 0.82 & 0.86                             \\
      misc          & cdk8\_koehler              & 10 & 0.44  & 0.65   & 0.38   & 0.00  & 0.90 & 0.68 & N/A \\
      misc          & galectin3\_extra           & 26 & 0.76  & 0.85   & 0.66   & 0.00  & 0.42 & 0.36 & 1.00                             \\
      misc          & hfaah                      & 24 & 0.72  & 0.66   & 0.50   & 0.00  & 0.75 & 0.60 & 0.98                             \\
      opls\_stress  & fxa\_set\_4                & 11 & 0.42  & 0.58   & 0.45   & 0.00  & 0.96 & 0.83 & 0.75                             \\
      opls\_stress  & fxa\_yoshikawa\_set        & 27 & 0.56  & 0.68   & 0.51   & 0.00  & 1.27 & 1.06 & 0.91                             \\
      opls\_stress  & hc\_bace\_1                & 7  & 0.67  & 0.68   & 0.52   & 0.00  & 0.65 & 0.53 & 1.00                             \\
      opls\_stress  & hc\_bace\_2                & 10 & 0.01  & -0.04  & 0.00   & 0.00  & 1.04 & 0.87 & 0.00                             \\
      waterset      & brd41\_ASH106              & 8  & 0.00  & 0.05   & 0.00   & 0.00  & 0.98 & 0.83 & 0.33                             \\
      waterset      & chk1                       & 13 & 0.62  & 0.76   & 0.61   & 0.00  & 0.95 & 0.74 & 0.94                             \\
      waterset      & hsp90\_kung                & 11 & 0.69  & 0.65   & 0.52   & 0.00  & 1.14 & 0.92 & 1.00                             \\
      waterset      & scyt\_dehyd                & 7  & 0.77  & 0.75   & 0.62   & 0.00  & 1.20 & 0.95 & 0.92                             \\
      waterset      & taf12                      & 8  & 0.83  & 0.74   & 0.64   & 0.00  & 0.34 & 0.32 & 1.00                             \\
      \bottomrule
    \end{tabular}
    \begin{flushleft}
    Here, \# denotes the number of ligands calculated in this work, while $\rho$ and $\tau$ denote Spearman's $\rho$ and Kendall's $\tau$, respectively.
    Offset, MAE, and RMSE are reported in \SI{}{\kcal\per\mole}.
    The FEP+ metrics and signed offsets are reproduced from the published benchmark data without further offset adjustment.
    Nonzero offsets are reported as supplied rather than being recentered in our analysis.
    \end{flushleft}
  \end{table}

\FloatBarrier
\subsection{Uncertainty in benchmark comparisons}
\label{sec:benchmark_uncertainty}

To quantify variation across benchmark targets, we computed pointwise 95\% confidence intervals using a paired cluster bootstrap.
The 43 benchmark systems were assigned to 30 target clusters, keeping systems that share a protein target together.
The shared-target clusters comprised BACE1 (six systems), HSP90 (five), and MCL1, CDK8, P2Y1, and factor Xa (two each).
The remaining 24 systems formed separate clusters.
For each of 100000 resamples, we drew 30 clusters with replacement and retained all systems in each selected cluster.
The same cluster draw was used for all methods, and the macro-average retained equal weight per benchmark system.
We used the 2.5th and 97.5th percentiles of the bootstrap distributions as interval bounds, with the PCG64 random-number generator and seed 20261029.
Resampling the fixed per-system metrics quantifies variation across targets without separately modeling experimental or trajectory uncertainty.
Tables~\ref{tab:benchmark_macro_ci} and~\ref{tab:benchmark_paired_ci} report the macro-averages and paired differences, respectively.
These intervals are pointwise and were not adjusted for multiple comparisons.

\begin{table}[htbp]
\centering
\caption{
Macro-averaged benchmark metrics and pointwise 95\% confidence intervals from the paired target-cluster bootstrap.
Each cell gives the estimate followed by its interval in brackets.
Offset-adjusted MAE and RMSE are reported in \SI{}{\kcal\per\mole}.
}
\label{tab:benchmark_macro_ci}
\begingroup
\small
\setlength{\tabcolsep}{4pt}
\begin{tabular}{lcccc}
\toprule
Method & Spearman's $\rho$ & Kendall's $\tau$ & \makecell{Offset-adjusted\\MAE} & \makecell{Offset-adjusted\\RMSE} \\
\midrule
\makecell[l]{Felis ABFE\\ByteFF25-AM1-BCC\\\SI{3}{\nano\second}$\times$3} & \makecell{0.658\\$[0.604, 0.712]$} & \makecell{0.510\\$[0.456, 0.567]$} & \makecell{0.927\\$[0.829, 1.032]$} & \makecell{1.178\\$[1.047, 1.319]$} \\
\makecell[l]{Felis ABFE\\ByteFF25-AM1-BCC\\\SI{5}{\nano\second}$\times$3} & \makecell{0.676\\$[0.619, 0.738]$} & \makecell{0.531\\$[0.467, 0.602]$} & \makecell{0.889\\$[0.795, 0.987]$} & \makecell{1.122\\$[1.001, 1.254]$} \\
\makecell[l]{FEP+ RBFE\\\SI{20}{\nano\second}} & \makecell{0.665\\$[0.616, 0.726]$} & \makecell{0.516\\$[0.471, 0.574]$} & \makecell{0.677\\$[0.613, 0.733]$} & \makecell{0.835\\$[0.759, 0.900]$} \\
\makecell[l]{Felis ABFE\\ByteFF25-ABCG2\\\SI{5}{\nano\second}$\times$3} & \makecell{0.630\\$[0.546, 0.715]$} & \makecell{0.497\\$[0.421, 0.577]$} & \makecell{0.896\\$[0.793, 1.003]$} & \makecell{1.140\\$[1.020, 1.259]$} \\
\bottomrule
\end{tabular}
\endgroup
\end{table}

\begin{table}[htbp]
\centering
\caption{
Paired differences in macro-averaged benchmark metrics with pointwise 95\% bootstrap confidence intervals.
Differences are calculated as the first method minus the second.
Positive differences favor the first method for rank correlations; negative differences favor it for offset-adjusted errors.
Offset-adjusted error differences are reported in \SI{}{\kcal\per\mole}.
}
\label{tab:benchmark_paired_ci}
\begingroup
\small
\setlength{\tabcolsep}{5pt}
\begin{tabular}{llrr}
\toprule
Comparison & Metric & Difference & 95\% confidence interval \\
\midrule
AM1-BCC: \SI{5}{\nano\second} minus \SI{3}{\nano\second} & Spearman's $\rho$ & $+0.019$ & $[-0.002, +0.044]$ \\
 & Kendall's $\tau$ & $+0.021$ & $[+0.001, +0.045]$ \\
 & MAE & $-0.039$ & $[-0.059, -0.020]$ \\
 & RMSE & $-0.057$ & $[-0.080, -0.031]$ \\
\midrule
Felis ABFE (\SI{5}{\nano\second}) minus FEP+ RBFE & Spearman's $\rho$ & $+0.012$ & $[-0.062, +0.084]$ \\
 & Kendall's $\tau$ & $+0.014$ & $[-0.055, +0.089]$ \\
 & MAE & $+0.212$ & $[+0.109, +0.328]$ \\
 & RMSE & $+0.286$ & $[+0.158, +0.439]$ \\
\midrule
AM1-BCC minus ABCG2 (\SI{5}{\nano\second}) & Spearman's $\rho$ & $+0.047$ & $[-0.015, +0.104]$ \\
 & Kendall's $\tau$ & $+0.033$ & $[-0.013, +0.074]$ \\
 & MAE & $-0.007$ & $[-0.087, +0.069]$ \\
 & RMSE & $-0.018$ & $[-0.121, +0.090]$ \\
\bottomrule
\end{tabular}
\endgroup
\end{table}
\FloatBarrier

\subsection{Benchmark-wide evaluation of RelocationMC}
\label{sec:rmc_benchmark_comparison}

\begin{figure}[p]
\centering
\includegraphics[width=\linewidth,height=0.86\textheight,keepaspectratio]{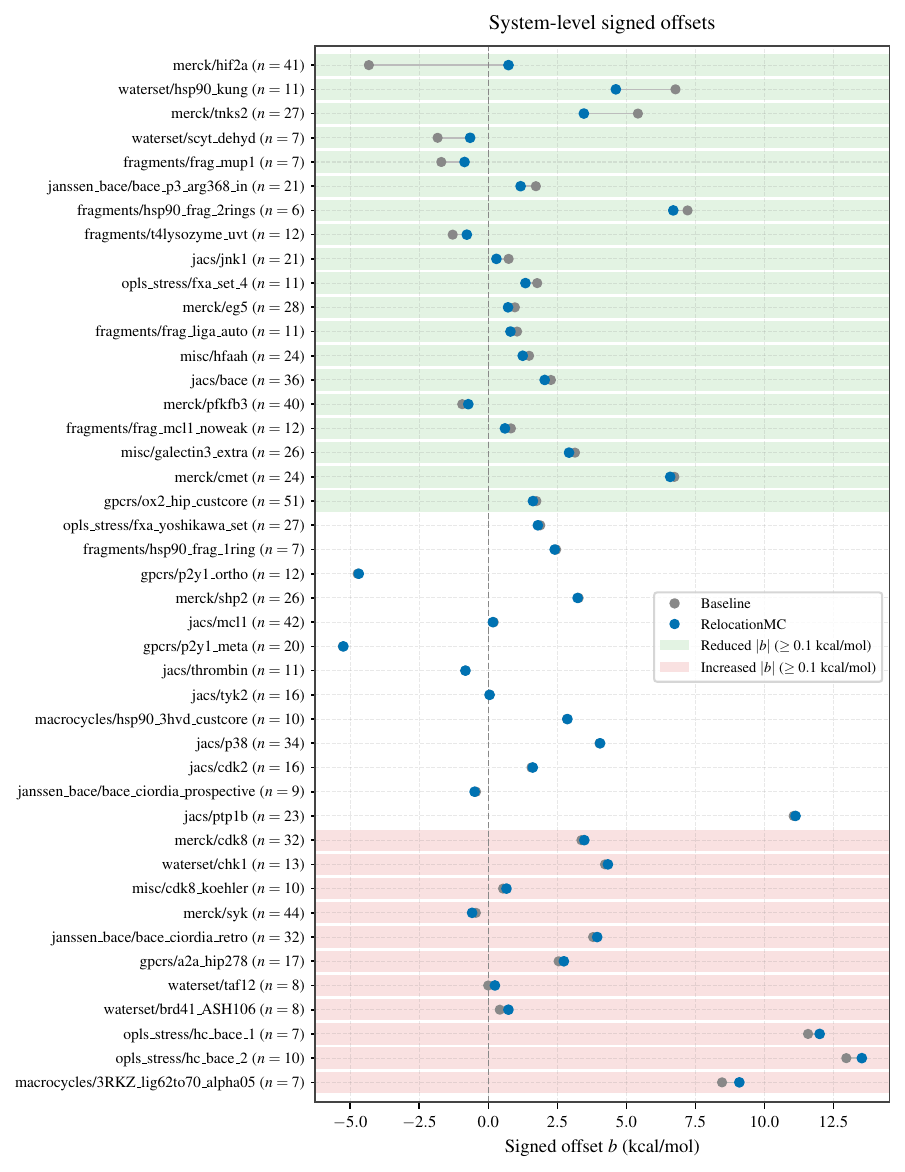}
\caption{
Signed system-level offsets for RelocationMC and baseline calculations across all 43 retained Schr\"odinger benchmark systems.
All simulation settings were matched apart from enabling RelocationMC.
The offset is $b=\langle\Delta G_{\mathrm{exp}}-\Delta G_{\mathrm{pred}}\rangle$, calculated from the mean of three independent predictions for each ligand.
Gray and blue points denote baseline and RelocationMC results, respectively; lines pair results for each system, and the dashed vertical line indicates zero offset.
Green shading marks a reduction in $|b|$ of at least \SI{0.1}{\kcal\per\mole}; red shading marks an increase of at least the same amount.
Unshaded rows indicate smaller changes, and $n$ denotes the number of ligand entries per system.
Color categories are descriptive and do not indicate statistical significance.
}
\label{fig:rmc_system_signed_offset}
\end{figure}

We compared the RelocationMC production results with baseline calculations for all 43 retained Schr\"odinger benchmark systems, using the same 857 ligand entries and experimental reference values.
Both protocols used \SI{5}{\nano\second} of production sampling per thermodynamic state, with identical force-field parameters, starting structures, and all other settings apart from enabling RelocationMC.
Predictions for both protocols were averaged over three independent calculations.
Figure~\ref{fig:rmc_system_signed_offset} shows the signed system-level offsets, with improvements defined by a reduction in their magnitudes.
RelocationMC reduced $|b|$ in 27 systems; 19 reductions and 11 increases were at least \SI{0.1}{\kcal\per\mole}, while the remaining 13 changes were smaller than this descriptive threshold.
The mean absolute offset decreased from \SI{3.23}{\kcal\per\mole} to \SI{2.97}{\kcal\per\mole}.
Macro-averaged rank correlations increased slightly, while offset-adjusted errors and the mean between-run standard deviation decreased (Table~\ref{tab:rmc_benchmark_comparison}).
These aggregate results support offset reduction in many systems without an overall deterioration in predictive performance, but do not establish improvement for every system or complete sampling convergence.

\begin{table}[htbp]
\centering
\caption{
Aggregate metrics for the RelocationMC and baseline calculations across 43 benchmark systems and 857 ligand entries.
All metrics except the between-run standard deviation are equally weighted averages of per-system metrics.
The sample standard deviation across three runs is averaged over ligand entries within each system and then equally over systems.
Offsets, errors, and standard deviations are reported in \si{\kcal\per\mole}.
}
\label{tab:rmc_benchmark_comparison}
\begin{tabular}{lrr}
\toprule
Metric & Baseline & RelocationMC \\
\midrule
Mean absolute offset $|b|$ & 3.23 & 2.97 \\
Spearman's $\rho$ & 0.666 & 0.676 \\
Kendall's $\tau$ & 0.520 & 0.531 \\
Offset-adjusted MAE & 0.938 & 0.889 \\
Offset-adjusted RMSE & 1.173 & 1.122 \\
Mean between-run standard deviation & 0.610 & 0.553 \\
\bottomrule
\end{tabular}
\end{table}

\FloatBarrier

\subsection{Water occupancy with crystallographic waters}
\label{sec:scyt_dehyd_water_with_crw}

\begin{figure}[htb]
\centering
\includegraphics[width=0.86\textwidth]{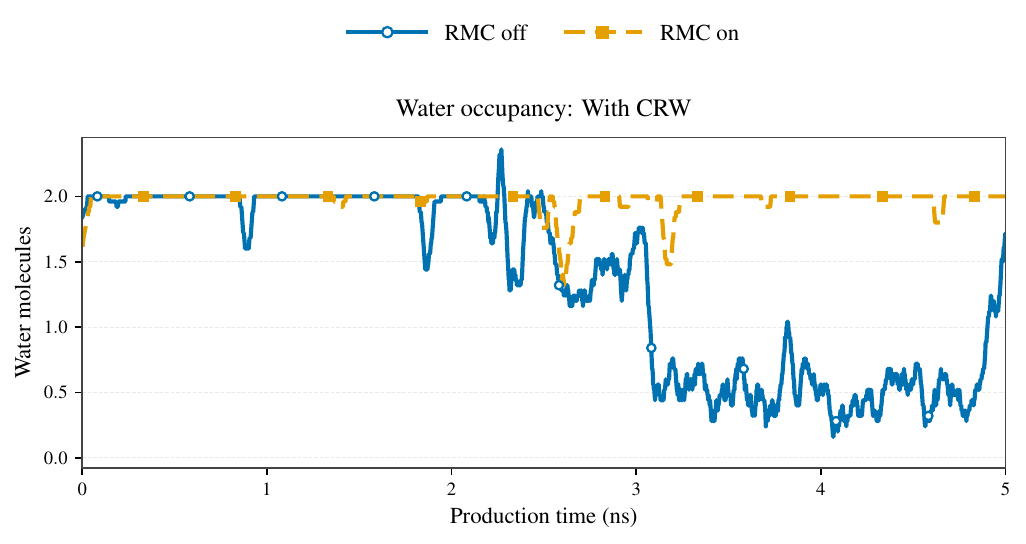}
\caption{
Binding-site water occupancy in scytalone dehydratase simulations initialized with crystallographic waters.
Each curve represents one fully coupled physical-state trajectory of ligand 3d.
Curves show the centered 25-frame rolling mean of the number of water molecules within \SI{3.5}{\angstrom} of the monitored site during the complete \SI{5}{\nano\second} production trajectory.
The site is centered on the minimum-image geometric center of the ligand nitrogen and the NE2 atoms of His79 and His104.
Blue circles and orange squares indicate simulations with RelocationMC disabled and enabled, respectively.
The corresponding simulations initialized without crystallographic waters are shown in Fig.~\ref{fig:scyt_dehyd_water_case_study}B.
}
\label{fig:scyt_dehyd_water_with_crw}
\end{figure}

The scytalone dehydratase trajectories initialized with crystallographic waters complement the initially dry-cavity simulations in Fig.~\ref{fig:scyt_dehyd_water_case_study}B.
The standalone water-occupancy trajectories in Fig.~\ref{fig:scyt_dehyd_water_case_study}B and Fig.~\ref{fig:scyt_dehyd_water_with_crw} were run separately from the HREMD ABFE production calculations.
These occupancy diagnostics used $K=16384$ candidate poses without crystallographic waters and $K=100$ with crystallographic waters.
The RelocationMC-enabled ABFE production calculations used $K=16384$ for both initial water configurations.
The occupancy diagnostics used relocation radii of \SI{0.843250}{\nano\meter} and \SI{0.886112}{\nano\meter} without and with crystallographic waters, respectively.
Relocation attempts occurred every \num{1000} integration steps (\SI{2}{\pico\second}) in both diagnostic runs.
Because the diagnostic candidate-pool sizes and relocation radii differed, the occupancy comparison does not isolate the effect of the initial water configuration under matched sampler settings.
With RelocationMC enabled, the occupancy remained near two water molecules, whereas without RelocationMC it decreased sharply after approximately \SI{3.1}{\nano\second} (Fig.~\ref{fig:scyt_dehyd_water_with_crw}).
Together with the initially dry trajectories, this comparison shows that the occupancy distributions remained sensitive to the protocol and starting water configuration over \SI{5}{\nano\second}.
The trajectories therefore do not establish complete equilibration of the buried hydration site.

\subsection{Discussion: the opls\_stress/hc\_bace\_2 set}
\begin{figure}[htb]
  \centering
  \includegraphics[width=0.8\textwidth]{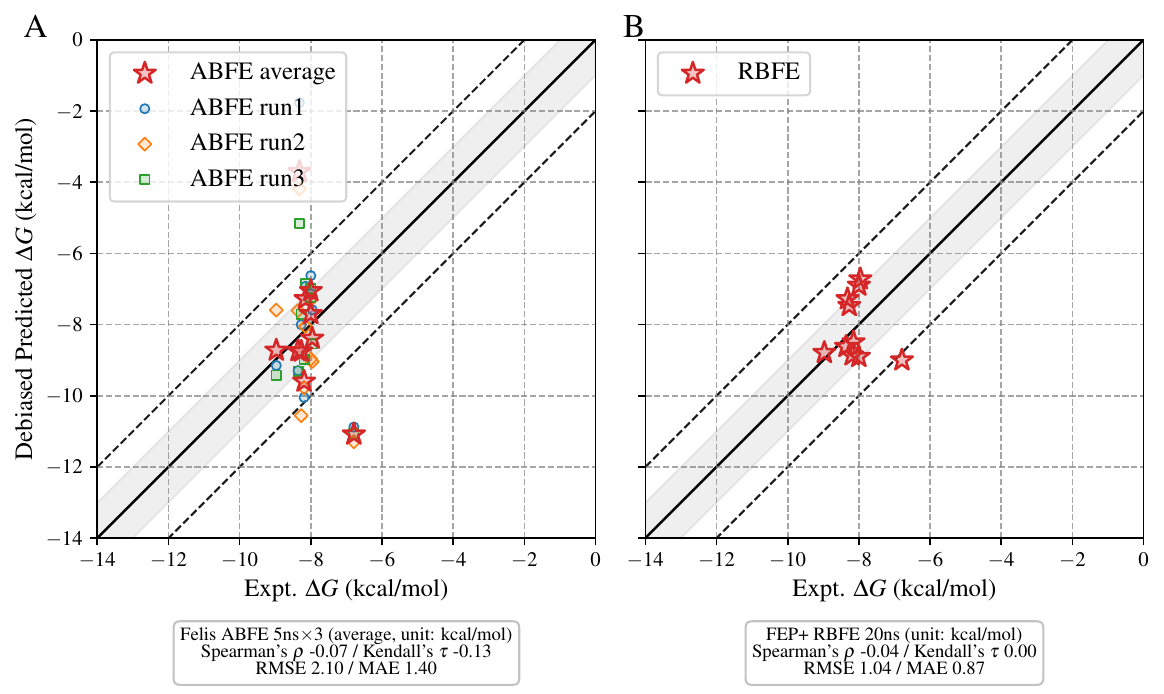}
  \caption{Comparison of Felis ABFE predictions using ByteFF25-AM1-BCC with FEP+ RBFE predictions~\cite{rossMaximalCurrentFEPBenchmark2023} for the opls\_stress/hc\_bace\_2 system.}
  \label{fig:hcbace2_scatter}
\end{figure}

For the opls\_stress/hc\_bace\_2 system, both Felis ABFE and FEP+ RBFE predictions failed to reproduce the experimental ranking (Fig.~\ref{fig:hcbace2_scatter}).
This system represents a challenging case with a very narrow dynamic range of experimental affinities.
The middle 8 of the 10 ligands span an energy interval of only \SI{0.4}{\kcal\per\mole}.
The difference between the most and least potent ligands is limited to \SI{2.18}{\kcal\per\mole}.
Notably, both methods failed to correctly rank the ligand pair with the lowest and highest binding affinity values.

\subsection{Discussion: KRAS(G12D) case study}
\label{sec:kras_details}

Convergence, neighboring-state overlap, and replica-mixing diagnostics for the ten-repeat MRTX1133 study are provided in \path{raw_data/kras_data/MRTX_1133_10repeats/}.
They cover both alchemical legs for all ten calculations at each of \SIlist{3;5;8;10}{\nano\second}.
The convergence plots show cumulative forward, cumulative backward, and independent-block free-energy estimates.
This diagnostic collection is limited to the MRTX1133 case study.

\begin{figure}[htb]
  \centering
  \includegraphics[width=0.86\textwidth]{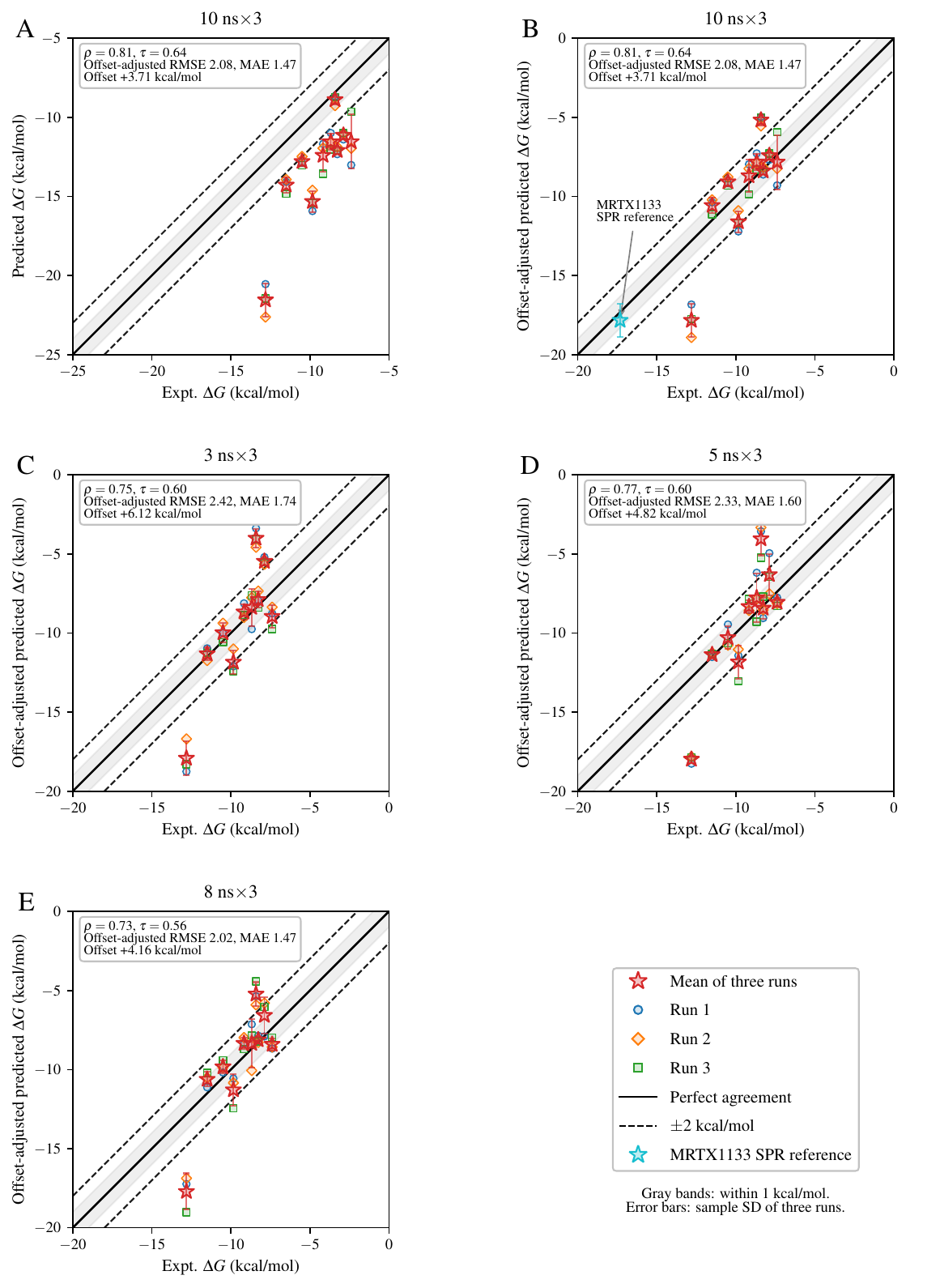}
  \caption{
  KRAS(G12D) ligand-level predictions before and after series-level offset adjustment.
  (A) Unadjusted \SI{10}{\nano\second} predictions.
  (B--E) Offset-adjusted predictions at \SIlist{10;3;5;8}{\nano\second}, respectively, using a separate signed offset for each duration.
  Stars show the mean of three postprocessed runs, and smaller symbols show individual runs.
  Vertical error bars show the sample standard deviation across the three runs.
  The solid diagonal indicates perfect agreement, gray bands denote deviations within \SI{1}{\kcal\per\mole}, and dashed lines denote deviations of \SI{2}{\kcal\per\mole}.
  The cyan star in panel B uses the nominal SPR-based reference for MRTX1133; this value is excluded from the calculation of offsets and statistical metrics.
  }
  \label{fig:kras_scatter}
\end{figure}

The unadjusted \SI{10}{\nano\second} ligand-level predictions are systematically more negative than the experimental references (Fig.~\ref{fig:kras_scatter}A).
The signed offsets at \SIlist{3;5;8;10}{\nano\second} are \SIlist{6.12;4.82;4.16;3.71}{\kcal\per\mole}, respectively.
For each duration, we add the corresponding signed offset to all three runs and their mean to obtain offset-adjusted predictions (Fig.~\ref{fig:kras_scatter}B--E).
Protein reorganization has been proposed as one explanation for series-level offsets in ABFE calculations from holo structures~\cite{chenEnhancingHitDiscovery2023}, but the present results do not identify its contribution.
After applying the signed offset at \SI{10}{\nano\second}, eight of the ten mean predictions lie within \SI{2}{\kcal\per\mole} of experiment.
MRTX1133 and MRTX119 remain outside this range.

The MRTX1133 reference of \SI{-12.81}{\kcal\per\mole} was taken from Uni-FEP~\cite{zouBreakingBarriersFEP2025} and retained for benchmark comparisons.
Wang et al. separately reported a cellular pERK IC\textsubscript{50} of \SI{2}{\nano\Molar} for MRTX1133~\cite{wangDiscoveryMRTX11332022}.
This assay-dependent inhibitory concentration is not treated as a dissociation constant or used to calculate the benchmark reference in our analysis.
The reported SPR estimate of $K_D\approx\SI{0.2}{\pico\Molar}$~\cite{wangDiscoveryMRTX11332022,hallin2022anti} gives a nominal binding free energy of \SI{-17.32}{\kcal\per\mole}.
The primary report identifies this affinity estimate as beyond the instrument's accurate determination range~\cite{wangDiscoveryMRTX11332022}.
The offset-adjusted \SI{10}{\nano\second} MRTX1133 prediction is \SI{-17.83}{\kcal\per\mole}, approximately \SI{0.51}{\kcal\per\mole} more negative than this nominal SPR reference (Fig.~\ref{fig:kras_scatter}B, cyan star).
The SPR value is shown as a qualitative alternative comparison rather than a precisely quantified experimental reference.
All offsets and statistical metrics retain the original Uni-FEP references, including the value for MRTX1133.

\begin{figure}[htb]
  \centering
  \includegraphics[width=\textwidth]{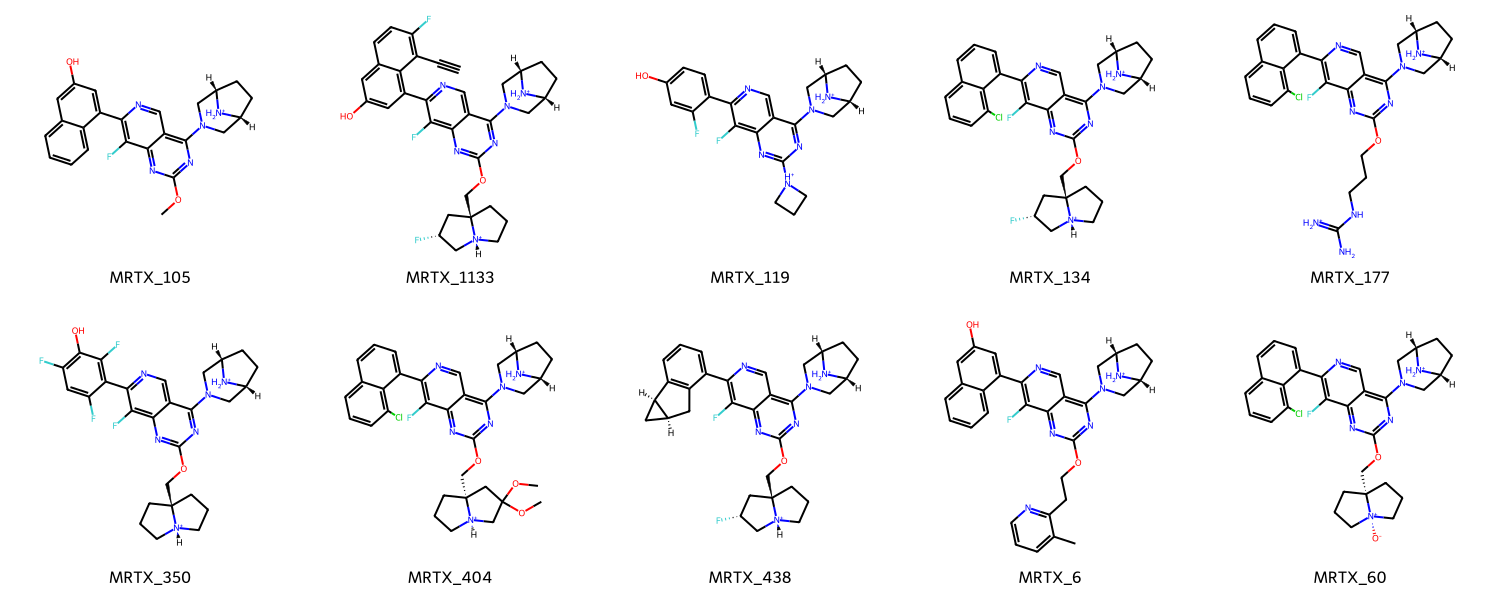}
  \caption{Ligand structures with explicit protonation states in the KRAS(G12D) benchmark series.}
  \label{fig:kras_ligands}
\end{figure}

\FloatBarrier
\section{Additional methodological details}

The initial ligand--solvent structure is placed in an orthorhombic periodic box with \SI{1.2}{\nano\meter} padding in all six directions ($\pm x, \pm y, \pm z$).
The protein--ligand structure is padded with \SI{1.0}{\nano\meter}.
No additional salt was added to the ligand--solvent or protein--ligand systems beyond the counterions required for charge neutrality.
This is consistent with the treatment used for relative binding free energy calculations without charge changes in~\citet{rossMaximalCurrentFEPBenchmark2023}.
This setup procedure is implemented in \texttt{ByteMol}, which is distributed with the Felis source code.
During structure preparation, we write a JSON metadata file that records atom categories (e.g., protein backbone, ligand, or water).

Both the PME real-space cutoff and the vdW cutoff are set to \SI{1.0}{\nano\meter}.
Production simulations use hydrogen mass repartitioning and a \SI{4}{\femto\second} integration time step.
For anchor-atom selection, we run four concurrent NPT trajectories at \SI{298.15}{\kelvin} and save snapshots every \SI{10}{\pico\second}.
Each trajectory is \SI{3}{\nano\second} long, and the first \SI{1}{\nano\second} is discarded as equilibration.
The retained portions are concatenated to provide a total of \SI{8}{\nano\second} for anchor selection.
For alchemical simulations, neighbor exchanges are attempted across the complete thermodynamic-state ladder every \SI{1}{\pico\second}.
Trajectory frames and restart checkpoints are written every \SI{0.5}{\nano\second}.
Felis uses one worker process and one \texttt{OpenMM} context per thermodynamic state.
Contiguous groups of workers are distributed as evenly as possible across the visible GPUs, with multiple contexts on each GPU executing concurrently through the CUDA Multi-Process Service.

\subsection{Alchemical schedules}

Felis defines separate default alchemical schedules for neutral and charged ligands in the solvent and protein--ligand legs.
The ligand--solvent leg uses 24 states for neutral ligands and 32 for charged ligands.
The corresponding protein--ligand leg uses 80 and 96 states.
Tables~\ref{tab:lambda_neutral_solvent}, \ref{tab:lambda_charged_solvent}, \ref{tab:lambda_neutral_protein}, and~\ref{tab:lambda_charged_protein} give the complete ordered default schedules for the ABFE v2 workflow in Felis v0.2.0.
Each row specifies the ligand charge factor $\lambda_{\mathrm{q},A}$, vdW factor $\lambda_{\mathrm{vdW},A}$, torsional factor $\lambda_{\mathrm{tors}}$, and Boresch restraint factor $\lambda_{\mathrm{rest}}$ at one thermodynamic state.
The complementary factors are $\lambda_{\mathrm{q},B}=1-\lambda_{\mathrm{q},A}$ and $\lambda_{\mathrm{vdW},B}=1-\lambda_{\mathrm{vdW},A}$.
For charged ligands, the complementary charge interpolation also transfers the removed ligand charge to the alchemical waters.

The ligand--solvent leg maintains $\lambda_{\mathrm{rest}}=0$ throughout, and its torsional factor follows a separate schedule from the vdW factor.
In the protein--ligand leg, restraints are introduced over the first eight states using factors of 0.00, 0.01, 0.03, 0.10, 0.30, 0.50, 0.75, and 1.00, then remain fully active.
The torsional factor in the protein--ligand leg equals $\lambda_{\mathrm{vdW},A}$ at every state.
The charge and vdW transformations overlap over a small range of states, as specified by the tables.
Each leg begins with a fully interacting ligand and ends with its alchemical charge, vdW, and scaled torsional terms switched off.

The same Boresch-style restraint force constants were used throughout the calculations.
For the harmonic distance and angle terms, the force constants are \SI{2}{\kcal\per\mole\per\angstrom^2} and \SI{80}{\kcal\per\mole\per\radian^2}, respectively.
For the cosine-based dihedral term, the force constant is \SI{80}{\kcal\per\mole}.

\begin{table}[p]
\centering
\caption{
Default alchemical schedule for the neutral ligand--solvent leg.
States are numbered from 1 and ordered down the left block, then down the right block.
}
\label{tab:lambda_neutral_solvent}
\begingroup
\footnotesize
\setlength{\tabcolsep}{4pt}
\begin{tabular}{rrrrr@{\hspace{1.5em}}rrrrr}
\toprule
State & $\lambda_{\mathrm{q},A}$ & $\lambda_{\mathrm{vdW},A}$ & $\lambda_{\mathrm{tors}}$ & $\lambda_{\mathrm{rest}}$ & State & $\lambda_{\mathrm{q},A}$ & $\lambda_{\mathrm{vdW},A}$ & $\lambda_{\mathrm{tors}}$ & $\lambda_{\mathrm{rest}}$ \\
\midrule
1 & 1.0000 & 1.0000 & 1.0000 & 0.0000 & 13 & 0.0000 & 0.5839 & 0.7084 & 0.0000 \\
2 & 0.8946 & 1.0000 & 1.0000 & 0.0000 & 14 & 0.0000 & 0.5337 & 0.6585 & 0.0000 \\
3 & 0.7875 & 1.0000 & 1.0000 & 0.0000 & 15 & 0.0000 & 0.4870 & 0.5954 & 0.0000 \\
4 & 0.6689 & 1.0000 & 0.9995 & 0.0000 & 16 & 0.0000 & 0.4462 & 0.5344 & 0.0000 \\
5 & 0.5354 & 1.0000 & 0.9768 & 0.0000 & 17 & 0.0000 & 0.4139 & 0.4323 & 0.0000 \\
6 & 0.3932 & 1.0000 & 0.9412 & 0.0000 & 18 & 0.0000 & 0.3880 & 0.3179 & 0.0000 \\
7 & 0.2376 & 0.9983 & 0.9082 & 0.0000 & 19 & 0.0000 & 0.3637 & 0.2454 & 0.0000 \\
8 & 0.0635 & 0.9626 & 0.8702 & 0.0000 & 20 & 0.0000 & 0.3411 & 0.1817 & 0.0000 \\
9 & 0.0017 & 0.8696 & 0.8318 & 0.0000 & 21 & 0.0000 & 0.3182 & 0.1278 & 0.0000 \\
10 & 0.0000 & 0.7723 & 0.7984 & 0.0000 & 22 & 0.0000 & 0.2855 & 0.0833 & 0.0000 \\
11 & 0.0000 & 0.6993 & 0.7699 & 0.0000 & 23 & 0.0000 & 0.2099 & 0.0466 & 0.0000 \\
12 & 0.0000 & 0.6382 & 0.7388 & 0.0000 & 24 & 0.0000 & 0.0000 & 0.0000 & 0.0000 \\
\bottomrule
\end{tabular}
\endgroup
\end{table}

\begin{table}[p]
\centering
\caption{
Default alchemical schedule for the charged ligand--solvent leg.
States are numbered from 1 and ordered down the left block, then down the right block.
}
\label{tab:lambda_charged_solvent}
\begingroup
\footnotesize
\setlength{\tabcolsep}{4pt}
\begin{tabular}{rrrrr@{\hspace{1.5em}}rrrrr}
\toprule
State & $\lambda_{\mathrm{q},A}$ & $\lambda_{\mathrm{vdW},A}$ & $\lambda_{\mathrm{tors}}$ & $\lambda_{\mathrm{rest}}$ & State & $\lambda_{\mathrm{q},A}$ & $\lambda_{\mathrm{vdW},A}$ & $\lambda_{\mathrm{tors}}$ & $\lambda_{\mathrm{rest}}$ \\
\midrule
1 & 1.0000 & 1.0000 & 1.0000 & 0.0000 & 17 & 0.0002 & 0.7921 & 0.5241 & 0.0000 \\
2 & 0.9368 & 1.0000 & 1.0000 & 0.0000 & 18 & 0.0000 & 0.7163 & 0.4948 & 0.0000 \\
3 & 0.8758 & 1.0000 & 1.0000 & 0.0000 & 19 & 0.0000 & 0.6587 & 0.4672 & 0.0000 \\
4 & 0.8144 & 1.0000 & 1.0000 & 0.0000 & 20 & 0.0000 & 0.6247 & 0.4468 & 0.0000 \\
5 & 0.7526 & 1.0000 & 0.9869 & 0.0000 & 21 & 0.0000 & 0.5937 & 0.4258 & 0.0000 \\
6 & 0.6901 & 1.0000 & 0.9550 & 0.0000 & 22 & 0.0000 & 0.5529 & 0.3950 & 0.0000 \\
7 & 0.6264 & 1.0000 & 0.9131 & 0.0000 & 23 & 0.0000 & 0.5066 & 0.3566 & 0.0000 \\
8 & 0.5611 & 1.0000 & 0.8737 & 0.0000 & 24 & 0.0000 & 0.4644 & 0.3181 & 0.0000 \\
9 & 0.4941 & 1.0000 & 0.8342 & 0.0000 & 25 & 0.0000 & 0.4255 & 0.2795 & 0.0000 \\
10 & 0.4246 & 1.0000 & 0.7948 & 0.0000 & 26 & 0.0000 & 0.3922 & 0.2414 & 0.0000 \\
11 & 0.3532 & 1.0000 & 0.7553 & 0.0000 & 27 & 0.0000 & 0.3639 & 0.2023 & 0.0000 \\
12 & 0.2807 & 1.0000 & 0.7158 & 0.0000 & 28 & 0.0000 & 0.3398 & 0.1636 & 0.0000 \\
13 & 0.2066 & 0.9998 & 0.6763 & 0.0000 & 29 & 0.0000 & 0.3171 & 0.1265 & 0.0000 \\
14 & 0.1306 & 0.9868 & 0.6367 & 0.0000 & 30 & 0.0000 & 0.2876 & 0.0865 & 0.0000 \\
15 & 0.0590 & 0.9438 & 0.5973 & 0.0000 & 31 & 0.0000 & 0.2328 & 0.0485 & 0.0000 \\
16 & 0.0097 & 0.8706 & 0.5591 & 0.0000 & 32 & 0.0000 & 0.0000 & 0.0000 & 0.0000 \\
\bottomrule
\end{tabular}
\endgroup
\end{table}

\begin{table}[p]
\centering
\caption{
Default alchemical schedule for the neutral protein--ligand leg.
States are numbered from 1 and ordered down the left block, then down the right block.
}
\label{tab:lambda_neutral_protein}
\begingroup
\footnotesize
\setlength{\tabcolsep}{4pt}
\begin{tabular}{rrrrr@{\hspace{1.5em}}rrrrr}
\toprule
State & $\lambda_{\mathrm{q},A}$ & $\lambda_{\mathrm{vdW},A}$ & $\lambda_{\mathrm{tors}}$ & $\lambda_{\mathrm{rest}}$ & State & $\lambda_{\mathrm{q},A}$ & $\lambda_{\mathrm{vdW},A}$ & $\lambda_{\mathrm{tors}}$ & $\lambda_{\mathrm{rest}}$ \\
\midrule
1 & 1.0000 & 1.0000 & 1.0000 & 0.0000 & 41 & 0.0000 & 0.8698 & 0.8698 & 1.0000 \\
2 & 1.0000 & 1.0000 & 1.0000 & 0.0100 & 42 & 0.0000 & 0.8458 & 0.8458 & 1.0000 \\
3 & 1.0000 & 1.0000 & 1.0000 & 0.0300 & 43 & 0.0000 & 0.8206 & 0.8206 & 1.0000 \\
4 & 1.0000 & 1.0000 & 1.0000 & 0.1000 & 44 & 0.0000 & 0.7951 & 0.7951 & 1.0000 \\
5 & 1.0000 & 1.0000 & 1.0000 & 0.3000 & 45 & 0.0000 & 0.7693 & 0.7693 & 1.0000 \\
6 & 1.0000 & 1.0000 & 1.0000 & 0.5000 & 46 & 0.0000 & 0.7442 & 0.7442 & 1.0000 \\
7 & 1.0000 & 1.0000 & 1.0000 & 0.7500 & 47 & 0.0000 & 0.7196 & 0.7196 & 1.0000 \\
8 & 0.9999 & 1.0000 & 1.0000 & 1.0000 & 48 & 0.0000 & 0.6953 & 0.6953 & 1.0000 \\
9 & 0.9938 & 1.0000 & 1.0000 & 1.0000 & 49 & 0.0000 & 0.6721 & 0.6721 & 1.0000 \\
10 & 0.9737 & 1.0000 & 1.0000 & 1.0000 & 50 & 0.0000 & 0.6500 & 0.6500 & 1.0000 \\
11 & 0.9329 & 1.0000 & 1.0000 & 1.0000 & 51 & 0.0000 & 0.6286 & 0.6286 & 1.0000 \\
12 & 0.8813 & 1.0000 & 1.0000 & 1.0000 & 52 & 0.0000 & 0.6074 & 0.6074 & 1.0000 \\
13 & 0.8267 & 1.0000 & 1.0000 & 1.0000 & 53 & 0.0000 & 0.5865 & 0.5865 & 1.0000 \\
14 & 0.7682 & 1.0000 & 1.0000 & 1.0000 & 54 & 0.0000 & 0.5664 & 0.5664 & 1.0000 \\
15 & 0.7089 & 1.0000 & 1.0000 & 1.0000 & 55 & 0.0000 & 0.5470 & 0.5470 & 1.0000 \\
16 & 0.6499 & 1.0000 & 1.0000 & 1.0000 & 56 & 0.0000 & 0.5273 & 0.5273 & 1.0000 \\
17 & 0.5893 & 1.0000 & 1.0000 & 1.0000 & 57 & 0.0000 & 0.5078 & 0.5078 & 1.0000 \\
18 & 0.5238 & 1.0000 & 1.0000 & 1.0000 & 58 & 0.0000 & 0.4887 & 0.4887 & 1.0000 \\
19 & 0.4533 & 1.0000 & 1.0000 & 1.0000 & 59 & 0.0000 & 0.4701 & 0.4701 & 1.0000 \\
20 & 0.3793 & 1.0000 & 1.0000 & 1.0000 & 60 & 0.0000 & 0.4519 & 0.4519 & 1.0000 \\
21 & 0.3027 & 1.0000 & 1.0000 & 1.0000 & 61 & 0.0000 & 0.4343 & 0.4343 & 1.0000 \\
22 & 0.2255 & 1.0000 & 1.0000 & 1.0000 & 62 & 0.0000 & 0.4174 & 0.4174 & 1.0000 \\
23 & 0.1551 & 1.0000 & 1.0000 & 1.0000 & 63 & 0.0000 & 0.4013 & 0.4013 & 1.0000 \\
24 & 0.1029 & 1.0000 & 1.0000 & 1.0000 & 64 & 0.0000 & 0.3857 & 0.3857 & 1.0000 \\
25 & 0.0705 & 1.0000 & 1.0000 & 1.0000 & 65 & 0.0000 & 0.3706 & 0.3706 & 1.0000 \\
26 & 0.0487 & 1.0000 & 1.0000 & 1.0000 & 66 & 0.0000 & 0.3561 & 0.3561 & 1.0000 \\
27 & 0.0318 & 1.0000 & 1.0000 & 1.0000 & 67 & 0.0000 & 0.3420 & 0.3420 & 1.0000 \\
28 & 0.0173 & 1.0000 & 1.0000 & 1.0000 & 68 & 0.0000 & 0.3285 & 0.3285 & 1.0000 \\
29 & 0.0064 & 1.0000 & 1.0000 & 1.0000 & 69 & 0.0000 & 0.3158 & 0.3158 & 1.0000 \\
30 & 0.0008 & 1.0000 & 1.0000 & 1.0000 & 70 & 0.0000 & 0.3034 & 0.3034 & 1.0000 \\
31 & 0.0000 & 0.9997 & 0.9997 & 1.0000 & 71 & 0.0000 & 0.2900 & 0.2900 & 1.0000 \\
32 & 0.0000 & 0.9966 & 0.9966 & 1.0000 & 72 & 0.0000 & 0.2745 & 0.2745 & 1.0000 \\
33 & 0.0000 & 0.9893 & 0.9893 & 1.0000 & 73 & 0.0000 & 0.2560 & 0.2560 & 1.0000 \\
34 & 0.0000 & 0.9798 & 0.9798 & 1.0000 & 74 & 0.0000 & 0.2345 & 0.2345 & 1.0000 \\
35 & 0.0000 & 0.9687 & 0.9687 & 1.0000 & 75 & 0.0000 & 0.2089 & 0.2089 & 1.0000 \\
36 & 0.0000 & 0.9559 & 0.9559 & 1.0000 & 76 & 0.0000 & 0.1782 & 0.1782 & 1.0000 \\
37 & 0.0000 & 0.9416 & 0.9416 & 1.0000 & 77 & 0.0000 & 0.1398 & 0.1398 & 1.0000 \\
38 & 0.0000 & 0.9259 & 0.9259 & 1.0000 & 78 & 0.0000 & 0.0946 & 0.0946 & 1.0000 \\
39 & 0.0000 & 0.9095 & 0.9095 & 1.0000 & 79 & 0.0000 & 0.0470 & 0.0470 & 1.0000 \\
40 & 0.0000 & 0.8911 & 0.8911 & 1.0000 & 80 & 0.0000 & 0.0000 & 0.0000 & 1.0000 \\
\bottomrule
\end{tabular}
\endgroup
\end{table}

\begin{table}[p]
\centering
\caption{
Default alchemical schedule for the charged protein--ligand leg.
States are numbered from 1 and ordered down the left block, then down the right block.
}
\label{tab:lambda_charged_protein}
\begingroup
\footnotesize
\setlength{\tabcolsep}{4pt}
\begin{tabular}{rrrrr@{\hspace{1.5em}}rrrrr}
\toprule
State & $\lambda_{\mathrm{q},A}$ & $\lambda_{\mathrm{vdW},A}$ & $\lambda_{\mathrm{tors}}$ & $\lambda_{\mathrm{rest}}$ & State & $\lambda_{\mathrm{q},A}$ & $\lambda_{\mathrm{vdW},A}$ & $\lambda_{\mathrm{tors}}$ & $\lambda_{\mathrm{rest}}$ \\
\midrule
1 & 1.0000 & 1.0000 & 1.0000 & 0.0000 & 49 & 0.0000 & 0.9344 & 0.9344 & 1.0000 \\
2 & 1.0000 & 1.0000 & 1.0000 & 0.0100 & 50 & 0.0000 & 0.9217 & 0.9217 & 1.0000 \\
3 & 1.0000 & 1.0000 & 1.0000 & 0.0300 & 51 & 0.0000 & 0.9083 & 0.9083 & 1.0000 \\
4 & 1.0000 & 1.0000 & 1.0000 & 0.1000 & 52 & 0.0000 & 0.8934 & 0.8934 & 1.0000 \\
5 & 1.0000 & 1.0000 & 1.0000 & 0.3000 & 53 & 0.0000 & 0.8766 & 0.8766 & 1.0000 \\
6 & 1.0000 & 1.0000 & 1.0000 & 0.5000 & 54 & 0.0000 & 0.8578 & 0.8578 & 1.0000 \\
7 & 0.9895 & 1.0000 & 1.0000 & 0.7500 & 55 & 0.0000 & 0.8373 & 0.8373 & 1.0000 \\
8 & 0.9608 & 1.0000 & 1.0000 & 1.0000 & 56 & 0.0000 & 0.8157 & 0.8157 & 1.0000 \\
9 & 0.9292 & 1.0000 & 1.0000 & 1.0000 & 57 & 0.0000 & 0.7930 & 0.7930 & 1.0000 \\
10 & 0.8981 & 1.0000 & 1.0000 & 1.0000 & 58 & 0.0000 & 0.7694 & 0.7694 & 1.0000 \\
11 & 0.8673 & 1.0000 & 1.0000 & 1.0000 & 59 & 0.0000 & 0.7448 & 0.7448 & 1.0000 \\
12 & 0.8347 & 1.0000 & 1.0000 & 1.0000 & 60 & 0.0000 & 0.7193 & 0.7193 & 1.0000 \\
13 & 0.8029 & 1.0000 & 1.0000 & 1.0000 & 61 & 0.0000 & 0.6934 & 0.6934 & 1.0000 \\
14 & 0.7712 & 1.0000 & 1.0000 & 1.0000 & 62 & 0.0000 & 0.6683 & 0.6683 & 1.0000 \\
15 & 0.7403 & 1.0000 & 1.0000 & 1.0000 & 63 & 0.0000 & 0.6449 & 0.6449 & 1.0000 \\
16 & 0.7087 & 1.0000 & 1.0000 & 1.0000 & 64 & 0.0000 & 0.6230 & 0.6230 & 1.0000 \\
17 & 0.6768 & 1.0000 & 1.0000 & 1.0000 & 65 & 0.0000 & 0.6016 & 0.6016 & 1.0000 \\
18 & 0.6451 & 1.0000 & 1.0000 & 1.0000 & 66 & 0.0000 & 0.5807 & 0.5807 & 1.0000 \\
19 & 0.6134 & 1.0000 & 1.0000 & 1.0000 & 67 & 0.0000 & 0.5610 & 0.5610 & 1.0000 \\
20 & 0.5818 & 1.0000 & 1.0000 & 1.0000 & 68 & 0.0000 & 0.5421 & 0.5421 & 1.0000 \\
21 & 0.5501 & 1.0000 & 1.0000 & 1.0000 & 69 & 0.0000 & 0.5232 & 0.5232 & 1.0000 \\
22 & 0.5185 & 1.0000 & 1.0000 & 1.0000 & 70 & 0.0000 & 0.5055 & 0.5055 & 1.0000 \\
23 & 0.4875 & 1.0000 & 1.0000 & 1.0000 & 71 & 0.0000 & 0.4885 & 0.4885 & 1.0000 \\
24 & 0.4567 & 1.0000 & 1.0000 & 1.0000 & 72 & 0.0000 & 0.4724 & 0.4724 & 1.0000 \\
25 & 0.4260 & 1.0000 & 1.0000 & 1.0000 & 73 & 0.0000 & 0.4571 & 0.4571 & 1.0000 \\
26 & 0.3950 & 1.0000 & 1.0000 & 1.0000 & 74 & 0.0000 & 0.4425 & 0.4425 & 1.0000 \\
27 & 0.3641 & 1.0000 & 1.0000 & 1.0000 & 75 & 0.0000 & 0.4285 & 0.4285 & 1.0000 \\
28 & 0.3349 & 1.0000 & 1.0000 & 1.0000 & 76 & 0.0000 & 0.4153 & 0.4153 & 1.0000 \\
29 & 0.3067 & 1.0000 & 1.0000 & 1.0000 & 77 & 0.0000 & 0.4028 & 0.4028 & 1.0000 \\
30 & 0.2787 & 1.0000 & 1.0000 & 1.0000 & 78 & 0.0000 & 0.3906 & 0.3906 & 1.0000 \\
31 & 0.2511 & 1.0000 & 1.0000 & 1.0000 & 79 & 0.0000 & 0.3786 & 0.3786 & 1.0000 \\
32 & 0.2230 & 1.0000 & 1.0000 & 1.0000 & 80 & 0.0000 & 0.3668 & 0.3668 & 1.0000 \\
33 & 0.1940 & 1.0000 & 1.0000 & 1.0000 & 81 & 0.0000 & 0.3554 & 0.3554 & 1.0000 \\
34 & 0.1638 & 1.0000 & 1.0000 & 1.0000 & 82 & 0.0000 & 0.3442 & 0.3442 & 1.0000 \\
35 & 0.1325 & 1.0000 & 1.0000 & 1.0000 & 83 & 0.0000 & 0.3329 & 0.3329 & 1.0000 \\
36 & 0.1007 & 1.0000 & 1.0000 & 1.0000 & 84 & 0.0000 & 0.3212 & 0.3212 & 1.0000 \\
37 & 0.0702 & 1.0000 & 1.0000 & 1.0000 & 85 & 0.0000 & 0.3086 & 0.3086 & 1.0000 \\
38 & 0.0450 & 1.0000 & 1.0000 & 1.0000 & 86 & 0.0000 & 0.2949 & 0.2949 & 1.0000 \\
39 & 0.0245 & 1.0000 & 1.0000 & 1.0000 & 87 & 0.0000 & 0.2794 & 0.2794 & 1.0000 \\
40 & 0.0094 & 1.0000 & 1.0000 & 1.0000 & 88 & 0.0000 & 0.2611 & 0.2611 & 1.0000 \\
41 & 0.0011 & 0.9997 & 0.9997 & 1.0000 & 89 & 0.0000 & 0.2393 & 0.2393 & 1.0000 \\
42 & 0.0000 & 0.9971 & 0.9971 & 1.0000 & 90 & 0.0000 & 0.2128 & 0.2128 & 1.0000 \\
43 & 0.0000 & 0.9919 & 0.9919 & 1.0000 & 91 & 0.0000 & 0.1830 & 0.1830 & 1.0000 \\
44 & 0.0000 & 0.9851 & 0.9851 & 1.0000 & 92 & 0.0000 & 0.1506 & 0.1506 & 1.0000 \\
45 & 0.0000 & 0.9771 & 0.9771 & 1.0000 & 93 & 0.0000 & 0.1156 & 0.1156 & 1.0000 \\
46 & 0.0000 & 0.9680 & 0.9680 & 1.0000 & 94 & 0.0000 & 0.0779 & 0.0779 & 1.0000 \\
47 & 0.0000 & 0.9578 & 0.9578 & 1.0000 & 95 & 0.0000 & 0.0382 & 0.0382 & 1.0000 \\
48 & 0.0000 & 0.9466 & 0.9466 & 1.0000 & 96 & 0.0000 & 0.0000 & 0.0000 & 1.0000 \\
\bottomrule
\end{tabular}
\endgroup
\end{table}

\FloatBarrier

\section{RelocationMC workflow and acceptance-ratio derivation}
\label{app:relocationmc-details}

\subsection{Paired-pool proposal}

Let $x$ denote the current configuration, and let the ligand-centered sphere partition the periodic box into inner and outer regions.
A relocation direction defines the current source region $S$ and the opposite target region $T$.
The direction is selected with probability $g_f$.
We write $\beta=(k_{\mathrm{B}}T_{\mathrm{therm}})^{-1}$, where $k_{\mathrm{B}}$ is the Boltzmann constant and $T_{\mathrm{therm}}$ is the temperature.
For each relocatable water $M$ in the source region, RelocationMC evaluates a source score $s_M$ and defines
\begin{equation}
Z_S(x)=\sum_{M\in S(x)}\exp(\beta s_M).
\end{equation}
The source water $D$ is then selected with probability
\begin{equation}
p(D\mid x,S)=\frac{\exp(\beta s_D)}{Z_S(x)}.
\end{equation}
The score includes the interaction of the water with its environment and the isolated-molecule reciprocal mesh contribution used by the finite-grid PME Hamiltonian.
This biased selection favors source waters with less favorable interactions.

RelocationMC next draws $K$ independent candidate anchors uniformly from the complete periodic box.
Each anchor is combined with an independent Haar-uniform rotation of the minimum-image conformation of $D$.
The candidates are classified as inner or outer after they are generated.
Candidates in the target region are assigned weights
\begin{equation}
w_i=\frac{\exp(-\beta u_i)}{h_0(i)},
\end{equation}
where $u_i$ is the interaction energy with the common environment after removal of $D$, and $h_0$ is the whole-box translation and Haar-orientation proposal density.
Because $h_0$ is constant for all candidates, it cancels from the normalized selection probabilities.
The implementation therefore uses $w_i=\exp(-\beta u_i)$.
One target-region candidate $C$ is selected with probability $w_C/W_f$, where
\begin{equation}
W_f=\sum_{i\in T}w_i.
\end{equation}

The reverse candidate pool is constructed by replacing the selected slot with the exact original pose of $D$ and retaining every other candidate.
This replacement is an involution because applying the same construction to the reverse move restores the forward pool.
The reverse-pool weight sum is denoted by $W_r$.
All target-region source scores are recomputed in the proposed configuration $y$, including $s_C$, to obtain
\begin{equation}
Z_T(y)=\sum_{M\in T(y)}\exp(\beta s_M).
\end{equation}
Empty eligible pools or nonfinite energies cause rejection without modifying the OpenMM context.

\subsection{Acceptance ratio}

The Metropolis--Hastings ratio for the proposed relocation is
\begin{equation}
r=\frac{\pi(y)q(y\rightarrow x)}{\pi(x)q(x\rightarrow y)},
\end{equation}
where $\pi$ is the equilibrium density and $q$ is the complete proposal probability.
The reverse direction is selected with probability $g_r$.
At fixed particle number in the NPT ensemble, the coordinate-dependent equilibrium ratio is
\begin{equation}
\frac{\pi(y)}{\pi(x)}=\exp(-\beta\Delta U),
\end{equation}
because a relocation leaves the box volume and particle identities unchanged.
The paired forward and reverse pools have identical generation densities.
Combining the direction, source-selection, and categorical candidate-selection probabilities gives
\begin{equation}
\frac{q(y\rightarrow x)}{q(x\rightarrow y)}
=
\frac{g_r}{g_f}
\frac{Z_S(x)}{Z_T(y)}
\exp\!\left[\beta(s_C-s_D)\right]
\frac{W_f}{W_r}
\exp\!\left[\beta(u_C-u_D)\right].
\end{equation}
Substitution into the Metropolis--Hastings ratio yields
\begin{align}
\log r ={}& \log(g_r/g_f)+\log Z_S(x)-\log Z_T(y) \nonumber\\
&+\beta(s_C-s_D)+\log W_f-\log W_r \nonumber\\
&-\beta\left[\Delta U-(u_C-u_D)\right].
\end{align}
The move is accepted with probability $\min(1,r)$.
Here, $\Delta U=U(y)-U(x)$, and $u_D$ and $u_C$ are the interaction energies of the original and proposed water poses with the common environment.
No explicit $V_T/V_S$ factor appears because both translations and orientations are drawn from the same whole-box proposal density before region classification.

\subsection{PME energy decomposition}

Candidate energies include direct-space and reciprocal-space PME interactions with the common environment.
Let $B$ denote the coordinates and charges of the common environment after removal of $D$.
The quadratic form of the reciprocal PME energy gives
\begin{equation}
U(B+C)-U(B+D)
=
u_C-u_D+\frac{1}{2}\left[r(C)-r(D)\right],
\end{equation}
where $r(P)$ is the isolated reciprocal mesh contribution of pose $P$.
The final term is a finite-grid correction rather than the coordinate-independent analytical Ewald self-energy.
RelocationMC includes this correction as the residual $\Delta U-(u_C-u_D)$ in the acceptance ratio.
The analytical Ewald self-energy, net-charge correction, and zero-valued intramolecular exception terms cancel separately.
The direct and reciprocal interactions for all candidates are evaluated in parallel on the GPU.
This exact-PME treatment replaces the reaction-field approximation used for water deletion in the original SwapMC implementation.

\subsection{Accepted-state update and Felis settings}

An accepted move changes only the coordinates and velocities of the selected water.
Let $Q$ be the proper best-fit rotation from the source water conformation to the accepted candidate conformation.
For site masses $m_i$ and source velocities $\mathbf{v}_i$, the center-of-mass velocity and updated site velocities are
\begin{align}
\mathbf{V}_{\mathrm{COM}} &= \frac{\sum_i m_i\mathbf{v}_i}{\sum_i m_i},\\
\mathbf{v}'_i &= \mathbf{V}_{\mathrm{COM}}+Q\left(\mathbf{v}_i-\mathbf{V}_{\mathrm{COM}}\right).
\end{align}
This update preserves the total linear momentum and the mass-weighted kinetic energy relative to the center of mass.
It replaces the velocity reset to zero used in SwapMC.
Rejected proposals leave all coordinates and velocities unchanged.

Felis applies RelocationMC only to the protein--ligand leg.
The production configuration uses $K=16384$ candidate poses and attempts one relocation every 1000 integration steps, corresponding to \SI{4}{\pico\second}.
The inner-sphere radius is
\begin{equation}
R_{\mathrm{in}}=
\frac{1}{2}\max_{i,j}\left|\mathbf{r}_i-\mathbf{r}_j\right|_{\mathrm{MI}}
+\SI{0.3}{\nano\meter},
\end{equation}
where $i$ and $j$ are ligand heavy atoms in the minimized bound pose and MI denotes the minimum-image distance.
The radius remains fixed during sampling, and the sphere is centered on the minimum-image geometric center of the ligand heavy atoms.
All relocatable waters have identical, lambda-invariant nonbonded parameters.
For charged ligands, charge-compensating alchemical waters are excluded because their nonbonded parameters vary along the alchemical pathway.
The current CUDA implementation requires an orthorhombic periodic box and zero-valued intramolecular nonbonded exceptions within each relocatable water.

\end{document}